\documentclass[superscriptaddress,twocolumn,prl]{revtex4-1}
\renewcommand{\vec}[1]{\mathbf{#1}}
\usepackage{amsfonts}
\usepackage{amsmath}    
\usepackage[normalem]{ulem}
\usepackage{bm}
\usepackage{graphicx}
\usepackage{titlesec}
\usepackage{tabularx}
\usepackage{color}
\usepackage{bbold}
\usepackage{upgreek,ulem,textcomp}

\usepackage[colorlinks=true,urlcolor =blue,linkcolor = black,citecolor = black]{hyperref}

\usepackage{amsfonts}
\usepackage{amsmath}    
\usepackage[normalem]{ulem}
\usepackage{bm}
\usepackage{graphicx}
\usepackage{titlesec}
\usepackage{tabularx}
\usepackage{color}
\usepackage{bbold}
\usepackage{upgreek,ulem,textcomp}
\usepackage{bbold}
\usepackage{comment}
\usepackage[british]{babel}
\usepackage{csquotes}
\usepackage{etoolbox}
\usepackage{amssymb}

\newcommand{\ket}[1]{\vert#1\rangle}

\newcommand{\Rmnum}[1]{\expandafter\@slowromancap\romannumeral #1@}

\begin{abstract}
Large-scale quantum networks require quantum memories featuring long-lived storage of non-classical light together with efficient, high-speed and reliable operation. The concurrent realization of these features is challenging due to inherent limitations of matter platforms and light-matter interaction protocols. Here, we propose an approach to overcome this obstacle, based on the implementation of the Autler-Townes-splitting (ATS) quantum-memory protocol on a Bose-Einstein condensate (BEC) platform. We demonstrate a proof-of-principle of this approach by storing short pulses of single-photon-level light as a collective spin-excitation in a rubidium BEC. For 20~ns long-pulses, we achieve an ultra-low-noise memory with an efficiency of $30\%$ and lifetime of 15~$\mu$s. The non-adiabatic character of the ATS protocol (leading to high-speed and low-noise operation) in combination with the intrinsically large atomic densities and ultra-low temperatures of the BEC platform (offering highly efficient and long-lived storage) opens up a new avenue towards high-performance quantum memories.

\end{abstract}

\begin{document}
\title{Storing short single-photon-level optical pulses in Bose-Einstein condensates for high-performance quantum memory}

\author{Erhan Saglamyurek}
\affiliation{Department of Physics, University of Alberta, Edmonton AB T6G 2E1, Canada}
\author{Taras Hrushevskyi}
\affiliation{Department of Physics, University of Alberta, Edmonton AB T6G 2E1, Canada}\author{Anindya Rastogi}
\affiliation{Department of Physics, University of Alberta, Edmonton AB T6G 2E1, Canada}
\author{Logan W. Cooke}
\affiliation{Department of Physics, University of Alberta, Edmonton AB T6G 2E1, Canada}
\author{Benjamin D. Smith}
\affiliation{Department of Physics, University of Alberta, Edmonton AB T6G 2E1, Canada}
\author{Lindsay J. LeBlanc}
\email{Corresponding authors: lindsay.leblanc@ualberta.ca, saglamyu@ualberta.ca }
\affiliation{Department of Physics, University of Alberta, Edmonton AB T6G 2E1, Canada}
\

\maketitle

Atomic systems are prime candidates for long-lived storage and on-demand retrieval of optical quantum states, due to the long coherence times of their optically accessible spin-states~\cite{Lvovsky2009b, Heshami2016c}. Several spin-based memory approaches  have been proposed and experimentally studied in a wide range of atomic media~\cite{Heshami2016c}, including warm~\cite{Phillips2001b} and cold atomic gases~\cite{Liu2001a}, rare-earth-ion doped solids~\cite{Turukhin2002a}, and single atoms in optical cavities~\cite{Boozer2007, Wilk2007} relying on various storage protocols, such as electromagnetically-induced-transparency (EIT)~\cite{Fleischhauer2000b}, off-resonant Raman~\cite{Nunn2007, Gorshkov2007}, and photon-echo~\cite{Moiseev2001, Afzelius2009a} techniques. To date, no combination of a platform and a protocol has been agreed upon as the ideal practical memory that concurrently features long lifetime~\cite{Dudin2013, Heinze2013}, efficient ~\cite{Hedges2010, Hosseini2011c, Hsiao2018a}, fast~\cite{Reim2010b, Guo2018} and reliable operation~\cite{Gundogan2015, Ding2015a, Vernaz-Gris2018, Wang2019b}, although these features have been demonstrated either individually or in pairs. Here, we introduce the Autler-Townes splitting (ATS) quantum memory protocol~\cite{Saglamyurek2018a} on a Bose-Einstein condensate (BEC) platform towards overcoming this obstacle. 

Bose-Einstein condensates (BECs) of alkali atoms were among the first-proposed light-storage platforms~\cite{VestergaardHau1999a, Dutton2004}, since a BEC's ultralow temperature inhibits thermal diffusion and thereby offers long-term storage~\cite{Zhang2009}. In addition, a BEC's large atomic density allows for strong light-matter coupling without optical cavities or particularly large atom numbers, leading to high-efficiency and high-speed quantum memory. Despite these intrinsic advantages, there have been only a few experimental studies exploring BECs for quantum memory. Early experiments focused on the long-lived storage and coherent manipulation of optical information in the classical domain~\cite{VestergaardHau1999a, Ginsberg2007, Zhang2009}, while more recent demonstrations tested the quantum nature of these processes~\cite{Lettner2011, Riedl2012}. All of these experiments used the EIT memory protocol~\cite{Fleischhauer2000b, Phillips2001b,Turukhin2002a, Lvovsky2009b, Dudin2013, Heinze2013, Heshami2016c, Hsiao2018a, Vernaz-Gris2018, Wang2019b}, which is favorable for efficient storage of long light-pulses but not well-suited to the short-pulse/large bandwidth storage regime~\cite{Gorshkov2007, Rastogi2019} due to this protocol's adiabatic nature. Moreover, the large optical densities and control-field powers required for a broadband EIT memory increase the impact of photonic noise processes, making reliable operation in the quantum regime difficult~\cite{Lauk2013, Geng2014a, Saglamyurek2019c}.

The ATS protocol~\cite{Saglamyurek2018a} overcomes these protocol-related limitations. In contrast to the EIT scheme,  the non-adiabatic (fast) character of this method allows for optimal storage of short light-pulses with substantially reduced technical demand and complexity~\cite{Rastogi2019}, and exceptional robustness to many noise processes~\cite{Saglamyurek2019c}. In this article, we present a proof-of-concept experimental implementation of the ATS protocol with a BEC to explore the unique advantages of this protocol-platform combination for a high-performance quantum memory. We demonstrate  efficient and ultra-low-noise storage of single-photon-level light pulses that are one-to-two orders of magnitude shorter than those reported in EIT based BEC-memories. We also show that  ATS based-storage in a BEC platform significantly outperforms its implementations in laser-cooled atoms.

\begin{figure}
\begin{center}
\includegraphics [width=0.47\textwidth]{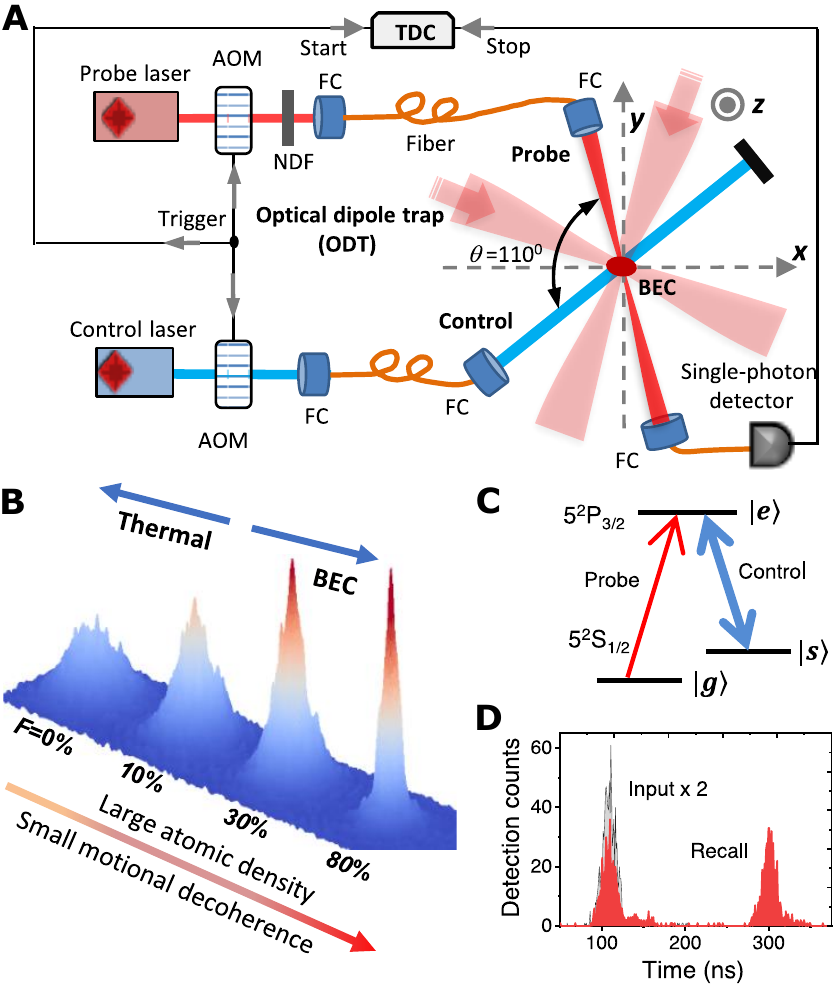}
\caption{\textbf{Demonstration of ATS memory in BEC.}~(\textbf{A})~Schematic of experimental setup. AOM: Acousto-optic modulator; NDF: Neutral density filter; FC: Fiber coupler; TDC: Time-to-digital convertor; ODT: Optical dipole trap. \textbf(\textbf{B}) Preparation of BEC, represented by velocity distributions for thermal, mixed, and condensated clouds (after $20$ ms of free expansion) at different stages of ODT-evaporation with temperatures of $T$ and BEC fractions of $\mathcal{F}_{\rm BEC}$. 
(\textbf{C}) $\Lambda$-system on the D2 transition of $^{87}$Rb.
(\textbf{D}) Storage of 20 ns-long probe pulses at the single photon level with an input-mean photon number of $\overline{n}_{\rm in}= 1$. The measured memory efficiency is $30\%$ under the conditions of $T=340$~nK and $\mathcal{F}_{\rm BEC} = 15\%$.
} 
\label{fig:setup}
\end{center}
\end{figure}

In our experiments, a BEC of $^{87}$Rb atoms is prepared using standard laser- and evaporative-cooling techniques~\cite{Lin2009}, with the resulting ultracold atoms held in an optical dipole trap (ODT) (Fig.~\ref{fig:setup}(\textbf{A}),(\textbf{B}) and Methods for details). By reducing the depth of the ODT, evaporative cooling drives the temperature of the atoms below the critical temperature $T_{\rm c}\approx0.5~\mu$K at which Bose-Einstein condensation begins (Fig.~\ref{fig:setup}($\textbf{B}$)). The fraction of condensed atoms ($\mathcal{F}_{\rm BEC}$) increases with further cooling, resulting in nearly pure BECs at $\mathcal{F}_{\rm BEC}\approx0.8$ and $T=280$ nK with the atom number of $N\approx10^5$ and characteristic spatial extent (Thomas-Fermi diameter) of $R_{\rm TF}\approx10~\mu\rm m$. Using the trap depth as a control, we study memory operation above and below the transition temperature.

The ATS protocol is implemented using a $\Lambda$-type three-level configuration within the ``D2'' transition of Rb atoms by addressing an excited level ($\ket{F^\prime = 2}\equiv\ket{e}$) and two ground hyperfine levels ($\ket{F= 1}\equiv\ket{g}$ and $\ket{F= 2}\equiv\ket{s}$ (Fig.~\ref{fig:setup}({\bf C})). In the storage (writing) stage, optical coherence from a weak ``probe'' pulse (resonant with $\ket{g}\rightarrow\ket{e}$) is transferred into collective excitations between the ground levels (spin-wave mode) via a strong ``control'' field  (coupled to the $\ket{s}\rightarrow\ket{e}$) with a pulse area of $2\pi$. By reapplying the control pulse after an adjustable storage time (read-out stage), the coherence from the spin-wave is mapped back to the optical mode, resulting in reemission as an output probe, as demonstrated in~Fig.~\ref{fig:setup}({\bf D})  

In our demonstrations, we use single-photon-level probe pulses with $\tau_{\rm p}=20$~ns duration (at full-width-half maximum of their Gaussian temporal profile), which is shorter than the natural lifetime of the ground-to-excited-level coherence  [$\tau_{\rm eg}=1/(2\pi\gamma_{\rm eg})=54\rm~{ns}$] of the Rb D2 line. Limited by our setup's focusing ability, the probe beam diameter (at $1/e^2$) is reduced to $R_{\rm p}\approx25~\mu\rm{m}$, but is still larger than the diameter of the atomic cloud $R_{\rm a}\approx~10~\mu\rm{m}$ for the lowest temperatures. To alleviate this size mismatch, we release the atomic cloud from the trap and allow 3.5~ms of free expansion before the storage-and-recall process, resulting in a cloud diameter comparable to that of the probe. The storage-and-recall process is then achieved using the write and read-out control fields with the same temporal profile of the probe pulses, but in a spatial mode oriented by $\theta=110^{\circ}$ away from the probe beam, as depicted in Fig.~\ref{fig:setup}({\bf A}).  The retrieved probe signal is detected via time-resolved photon-counting measurements using a single-photon detector (SPD) and a time-to-digital converter (TDC), allowing the evaluation of memory performance by recording detection-vs.-time histograms, as detailed in Methods.

\begin{figure*}
\begin{center}
\includegraphics[width = 170mm]{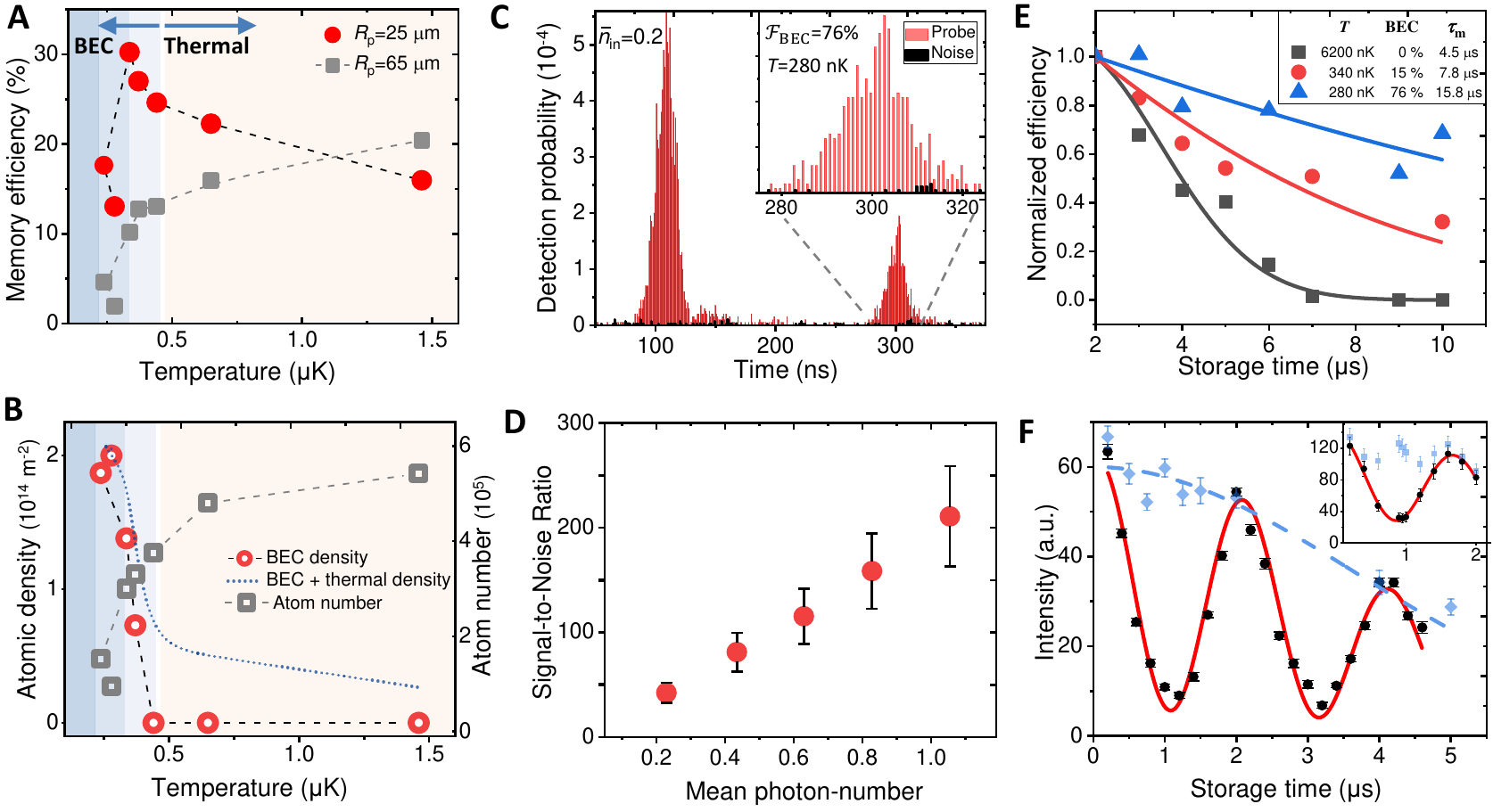}
\caption{\textbf{Experimental results for ATS-BEC memory.}~
\textbf{(A) Efficiency}. Measured efficiency is $\eta_{\rm m} = (p_{\rm s} - p_{\rm n}) / p_{\rm in}$, where $p_{\rm s}$, $p_{\rm n}$, and $p_{\rm in}$ are the detection probabilities for recalled probe, noise, and input probe, respectively, with $\overline{n}_{\rm in}= 1$ and $p_{\rm n} \ll p_{\rm s}$. 
\textbf{(B) Atom number and density}, estimated for 3.5~ms free- expansion time using $T,~\mathcal{F}_{\rm BEC}$ and ODT parameters, as detailed in Supplementary Information. The density refers to peak value for the cross-sectional density profile, obtained by the line integration of volume atomic density along the probe propagation direction. 
\textbf{(C) Low noise operation}. Measurement histograms in red (with probe) and black (without probe) show $p_{\rm s}$ and $p_{\rm n}$ respectively, after 200~ns storage in a nearly pure BEC for $\overline{n}_{\rm in}= 0.2$. The inset shows the time interval in which the retrieved photons are detected.
\textbf{(D) Signal-to-noise ratio vs input photon number}. ${\rm SNR}=(p_{\rm s}-p_{\rm n})/p_{\rm n}$ is determined for each mean photon number by measuring $p_{\rm s}$ for $\overline{n}_{\rm in} \neq 0$ and $p_{\rm n}$ for $\overline{n}_{\rm in} = 0$, during a 50~ns time window centered around the recall time.
\textbf{(E) Memory lifetime}. Variation of efficiency (normalized to its own maximum) with storage time for temperatures corresponding to thermal, mixed and nearly-pure BEC clouds, giving $1/e$ decay times of $\tau_{\rm m}=4.5~\mu\rm{s,}~7.8~\mu\rm{s~and}~15.8~\mu\rm{\rm s}$, respectively
\textbf{(F) Preservation of phase}. Retrieved probe intensity vs storage time with magnetic field off (blue diamonds) and on (black squares). The dashed blue and solid red are fits to functions involving memory decoherence (Eq.~\ref{eq:decoherence1} in Methods) and a product of the decoherence with a sinus, respectively. The visibility $V=(I_{\rm max}-I_{\rm min})/(I_{\rm max}+I_{\rm min})$, where $I_{\rm max}$ and $I_{\rm min}$ are the maximum and minimum intensities for zero storage time, yields $V=80 \pm 3~\%$ and $V=62 \pm 7 ~\%$ (inset, same axes) for $\overline{n}_{\rm in}\gg1$ and $\overline{n}_{\rm in}=1$.
}
\label{fig:single-photon}
\end{center}
\end{figure*}


We characterize the performance of the ATS-BEC memory at various temperatures, ranging from above the condensation temperature (with $\mathcal{F}_{\rm BEC}=0$) to well below the transition to BEC (where $\mathcal{F}_{\rm BEC}\rightarrow1$). First, memory efficiency $\eta_{\rm m}$ is measured at temperatures between $1.5~\mu$K to 280~nK (corresponding to thermal and nearly pure-BEC clouds, respectively) for an average input-probe-photon number of $\overline{n}_{\rm in}= 1$ and storage time of $\tau_{\rm s}=200~\rm{ns}$ (Fig.~\ref{fig:single-photon}). We observe that efficiency increases as the temperature is reduced (Fig.~\ref{fig:single-photon}({\bf A})), due to a significant increase in peak atomic density associated with BEC (Fig.~\ref{fig:single-photon}({\bf B})). When the BEC fraction is $\mathcal{F}_{\rm BEC}\approx15\%$ at $T\approx340~\rm nK$, the efficiency reaches its maximum $\eta_{\rm m}=(30.2\pm1.5)\%$. However, further evaporation leads to a reduction in efficiency with $\eta_{\rm m}=(13.0\pm0.9)\%$ for the nearly pure BEC at $T\approx280~\rm nK$ despite an additional increase in the peak density. We attribute the loss of efficiency to the limited ability to focus the probe beam onto a sufficiently small area at the center of the BEC, where the atomic density is largest. In such limits of $R_{\rm p} \gg R_{\rm a}$ or $R_{\rm p} \sim R_{\rm a}$ as in our demonstrations, the efficiency does not entirely follow the variation in peak atomic density (unfilled circles,  Fig.~\ref{fig:single-photon}({\bf B})). Instead, it is either partly or fully governed by the atom number, which inevitably decreases during the evaporative cooling (unfilled squares, Fig.~\ref{fig:single-photon}({\bf B})). To verify this conjecture, we repeat the efficiency measurements with larger beam size $R_{\rm p}\approx65~\mu\rm{m}$ (grey squares, Fig.~\ref{fig:single-photon}({\bf A})) and find that memory efficiency decreases monotonically with temperature over the entire range and does, indeed, track the variation in atom-number. This also shows that the size mismatch between the probe and BEC is the primary limitation to reaching high efficiencies in our setup: the free expansion alleviates this mismatch at the expense of an overall reduction in the atomic density and hence efficiency.


Next, we investigate the variation of memory lifetime $\tau_{\rm m}$ with respect to temperature $T$ in both the thermal and BEC regimes. We measure the efficiency of ATS memory as a function of storage time (from $\tau_{\rm s}=2 \rm~to~10~\mu \rm s$) at three different temperatures between $T=280$~nK and $6200$~nK 
and determine the lifetime (defined as storage time for which efficiency decreases to $1/e$ of its original value) for each $T$. As the temperature is lowered towards the BEC regime, memory lifetime increases significantly and reaches a maximum of $\tau_{\rm m}=15~\mu \rm s$ at $T=280$~nK for the nearly pure BEC, as shown in Fig.~\ref{fig:single-photon}({\bf E})). We attribute this observation to the combined effect of two spin-decoherence mechanisms:  thermally induced atomic-diffusion (given by cloud temperature), and magnetic dephasing (due to uncancelled ambient magnetic fields). Memory lifetime is mainly limited by thermal decoherence at relatively high temperatures ($T>3~\mu$K) in the thermal regime, while magnetic decoherence dominates in the BEC regime, where thermal motion is suppressed. Since thermal decoherence also increases with probe-control separation angle ($\theta$), inhibiting thermal diffusion in the BEC regime provides a flexibility to use large angles ($\theta>2^\circ$, as in our demonstrations), where low-noise operation can be realized.

\begin{figure*}
\begin{center}
\includegraphics[width = 150mm]{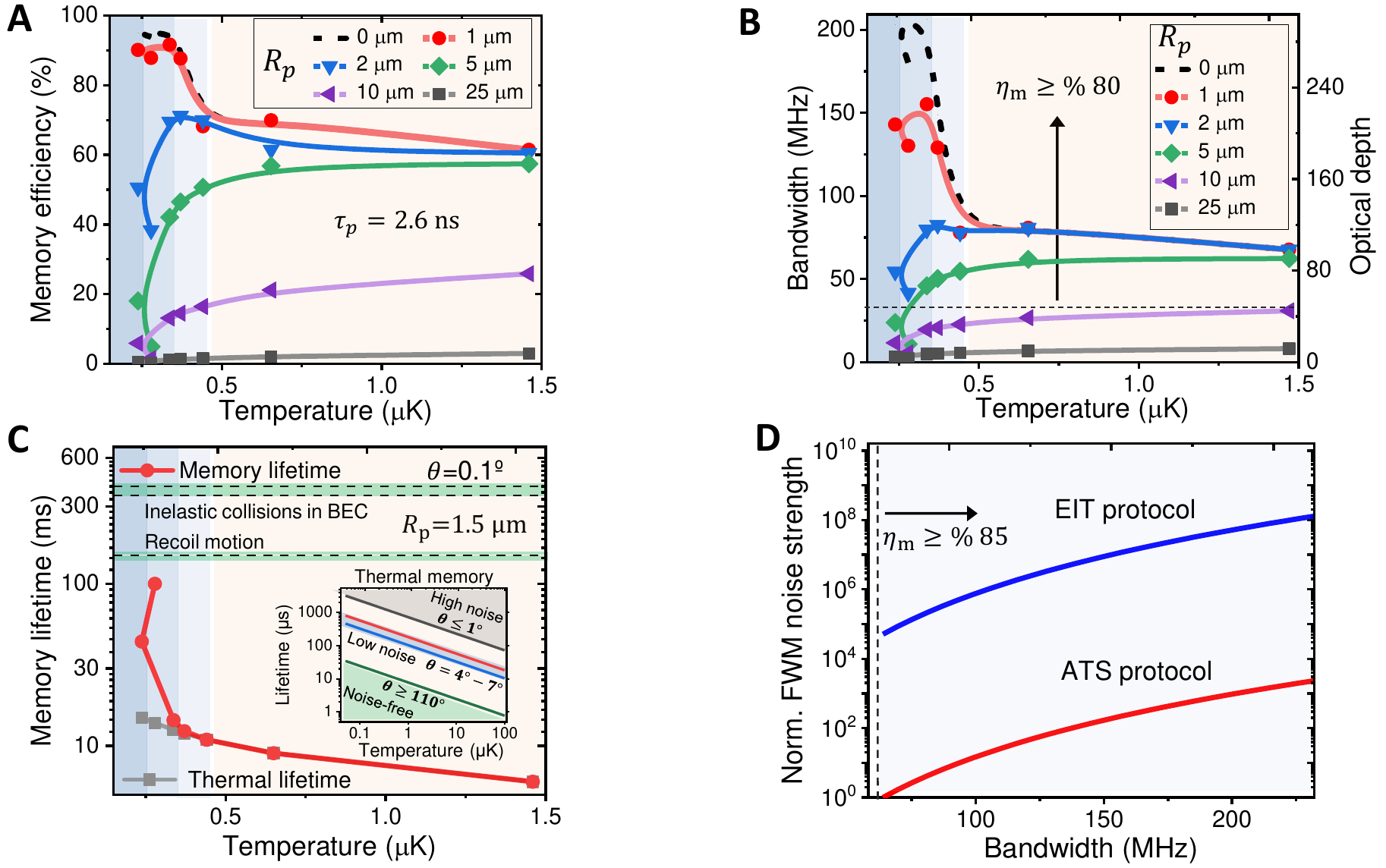} 
\caption{\textbf{Predicting ATS-BEC memory performance.} All calculations use experimentally measured atom numbers, T, $\mathcal{F}_{\rm BEC}$, and ODT frequencies. We assume that atoms remain trapped (no free-expansion) during storage and ATS memory is implemented in the backward recall scheme using the D1 transition of $^{87}$Rb.
\textbf{(A) Memory efficiency vs. temperature}, estimated for short storage times ($\tau_s\ll \tau_m$), memory bandwidth of $B=170$~MHz ($\tau_p=2.6$~ns), and probe-beam diameters $R_p=1-25~\mu\rm m$. The dashed black line refers to peak optical density. 
\textbf{(B) Bandwidth and optical depth vs. temperature}, estimated for an optimal ATS memory, where $d/2F=d\Gamma_{\rm eg}/4\pi B\approx4$. The horizontal dashed line indicates bandwidths and optical depths yielding efficiencies above $80\%$. 
\textbf {(C) Memory lifetime vs.\ temperature}, for a small probe-control separation angle ($\theta$), based on the combination of decoherence effects due to thermal motion, recoil momentum and inelastic two-body collisions, as detailed in Methods. The inset considers only the thermal memory lifetime $\tau_{\rm th}$ (Eq.~\ref{eq:dec} in Methods) for a wide range of temperatures and separation angles associated with different photonic noise levels.
\textbf{(D) FWM noise strength vs.\ bandwidth} calculated for optimal ATS and EIT memories, as detailed in Methods. The noise strength is normalized with respect to its minimum value corresponding to the smallest bandwidth of $10\Gamma_{\rm eg}/ 2\pi$. The vertical dashed line indicates bandwidths yielding efficiencies above $85\%$ 
}
\label{fig:performance}
\end{center}
\end{figure*}


To demonstrate simultaneous low-noise and broadband operation, we implement ATS memory in the nearly-pure BEC for several mean input-photon-numbers less than unity (Fig.~\ref{fig:single-photon}({\bf C})) shows measurements for $\overline{n}_{\rm in}\approx0.2$). We determine the signal-to-noise ratio ${\rm SNR}=(p_{\rm s}-p_{\rm n})/p_{\rm n}$ as a function of $\overline{n}_{\rm in}$ after $\tau_{\rm s}=200$ ns, where $p_{\rm s}$ and $p_{\rm n}$ are independently measured detection probabilities for retrieved probe and noise. We measure a background-noise probability of $p_{\rm n}=(6.6\pm1.5)\times10^{-5}$ (inset of Fig. \ref{fig:single-photon}({\bf C})), yielding an $\rm SNR\gtrsim100$ in much of the low mean-photon-number regime ($\overline{n}_{\rm in}<1$), as illustrated in Fig. \ref{fig:single-photon}({\bf D})). The SNR can be as high as $42\pm9$, even for mean photon-numbers as small as $\overline{n}_{\rm in}=0.22$, which is typical in quantum photonics applications. This SNR would yield an error probability of  $\mathcal{E}=1/\rm{SNR}=0.023\pm0.005$, if quantum states were encoded into these photons. Our additional characterisations show that the observed residual noise comes from scattered light leaking from both ODT and control beams, which can be almost entirely eliminated with simple technical upgrades. This also implies that there is no measurable noise contribution from any physical process linked to the memory operation, such as the four-wave mixing (FWM) noise, demonstrating the reliability of ATS-BEC memory for short pulses at the single-photon level.


Finally, we examine the phase-preserving character of photon storage process in our implementation (at near $T_{\rm c}$) by controlling the phase evolution of the stored spin-wave. We apply a weak DC magnetic field  to the ensemble such that $\ket{g}$ and $\ket{s}$ levels (forming spin-wave coherence) are split into the Zeeman sublevels with energy/frequency differences proportional to the strength of the field~\cite{Jenkins2006}. In the writing stage, we then map optical coherence onto different classes of spin-waves among these Zeeman levels with the proper selection of the magnetic field orientation, and polarizations of probe and control-field (see Supplementary Information for details and also Refs~\cite{Matsukevich2006, Wang2011, Farrera2018}). Since each class of spin-wave evolves with a different frequency, they acquire relative phase differences and thus interfere with one another, either constructively or destructively, resulting in the intensity of the recalled probe being modulated with storage time, as shown in Fig.~\ref{fig:single-photon}(\textbf{F}). We achieve an interference visibility of $V=62\%$ for small input photon-number $\overline{n}_{\rm in}=1$ (inset), while reaching up to $V=80\%$ for large mean photon numbers due to a better magnetic-field stability with single-shot measurements. These results demonstrate the phase-preserving nature of our memory, which is a key requirement for quantum information storage.


Looking beyond these proof-of-principle demonstrations, we find that the ATS-BEC approach is suitable for a high performance quantum memory, featuring the co-existence of highly efficient and long-lived storage together with broadband and low-noise operation. Particularly, the relaxed optical-depth demand of the ATS protocol in conjunction with the ultra-large optical densities of the BEC platform renders near-unity efficiencies at GHz storage bandwidths. This performance can be achieved in our system by sampling the dense region of BEC with a probe-beam diameter that is significantly smaller than the Thomas-Fermi diameter ($R_{\rm p}\ll R_{\rm a} \approx R_{\rm TF}$). Fig.~\ref{fig:performance}(\textbf{A}) shows the predicted memory efficiency with respect to temperature for pulses as short as $\tau_{\rm p}=2.6$~ns and smaller probe-beam diameters than used in our demonstrations (see Methods for details). An effective optical depth of $d\approx200$ is possible for $R_{\rm p}=1~\mu\rm m$ at $T=280$~nK ($\mathcal{F}_{\rm BEC} \approx 0.8$), allowing a near-optimal memory efficiency of $\eta_{\rm m }\ge90\%$. We also predict the acceptance bandwidth and optical depth for an optimal ATS memory with respect to temperature, as shown in Fig.~\ref{fig:performance}B, confirming the feasibility of bandwidths approaching $200$ MHz with efficiencies above $\eta_{\rm m }\ge90\%$. The same performance via EIT or off-Resonant Raman memory would require an optical depth of about $d=1000-1500$~\cite{Saglamyurek2018a, Rastogi2019}, which is hard to achieve even with typical BEC systems. 

An ATS-BEC memory can also reach long lifetimes from milliseconds to a second by reducing the impact of the three major spin-decoherence mechanisms: magnetic dephasing, thermal diffusion, and internal/external dynamics of BEC~\cite{Dutton2004}. First, magnetic dephasing can be eliminated using well-mastered techniques, including a high-degree isolation from the static and time-dependent magnetic-field noise, spin-echo dephasing/rephasing schemes, and the precise control over magnetic-insensitive Zeeman states~\cite{Zhao2008, Dudin2013}. Second, thermal diffusion, which is an increasing function of both $T$ and $\theta$, is already significantly reduced due to ultra-low temperatures in our evaporatively cooled system. Beyond that, as the thermal velocity is virtually zero in the nearly-pure BEC, BEC memory features much longer lifetimes than what is achievable with a purely thermal cloud even at ultracold temperatures and small $\theta=0.1^\circ$, as shown in Fig.~\ref{fig:performance}(\textbf{C}). Third, BEC-specific decoherence mechanisms are already effective only at long time scales over milliseconds, in part, because of the coherent matter-wave nature of BEC. Among these mechanisms, spatial decoherence (arising from atomic motions in the trap) can be coherently compensated using matter-wave interferometry techniques. However, recoil motion (an increasing function of $\theta$) and inelastic two-body collisions (proportional to atomic density) set an ultimate limit to memory lifetime. Considering the combined effect of these two mechanisms together with thermal diffusion, we predict that memory lifetime in our Rb-system can reach  one hundred milliseconds (Fig.~\ref{fig:performance}(\textbf{C})).

Low-noise operation is another essential requirement that is difficult to satisfy simultaneously with long lifetimes and large bandwidths. In particular, noise from control-leak and FWM processes can be minimized using a large-$\theta$ configuration, but this conflicts with the small-$\theta$ requirement for reduced thermal decoherence. In comparison to laser-cooled systems,  the small thermal-diffusion rates at ultracold temperatures ($T \lesssim 1~\mu$K) allow one to overcome the detriment of large-$\theta$ and thus provide a workable range of $\theta\approx 4-7^\circ$, where  low-noise operation is possible while retaining a millisecond lifetime (inset of Fig.~\ref{fig:performance}(\textbf{D})). However,  lifetimes of order one-hundred milliseconds still require a nearly-pure-BEC ensemble (at $T\rightarrow 0$) with a small $\theta$,  which prevents decoherence due to recoil motion. In this scenario ($\tau_{\rm m} \gg 10$ ms and $\theta < 0.2^\circ$), low-noise operation favours small optical depths and control powers, contrasting with the high demand on these resources for a broadband memory. In particular, FWM noise is a significant issue, due to its exponential and quadratic dependencies on optical depth and control power, respectively (as detailed in Methods)~\cite{Lauk2013, Geng2014a, Saglamyurek2019c}. With respect to the adiabatic protocols like EIT and Raman, the ATS protocol is very advantageous to reduce FWM noise in the broadband regime due to the substantially lower requirements for these resources. Fig. \ref{fig:performance}(\textbf{D}) compares the estimated relative strength of FWM noise vs.\ bandwidth between the optimal implementations of the EIT and ATS protocols in our Rb-system. This prediction shows that the probability of FWM noise with ATS memory is four to five orders of magnitude less compared to that of EIT memory, for storing a few nanoseconds-long pulses at near-unity efficiencies. 

In conclusion, we have experimentally demonstrated the non-adiabatic storage of single-photon-level light in a rubidium-87 BEC using the ATS protocol with a pulse duration that is 1-2 orders of magnitude shorter than those reported in previous BEC memories. Our proof-of-principle experiments and predictive analysis highlight the inherent advantages of the ATS protocol-BEC platform combination for a high-performance quantum memory, simultaneously featuring high efficiency, long lifetime, high-speed and low noise operation. In view of the recent technical progress with portable, miniaturized, and even space-based BEC experiments~\cite{Elliott2018}, we anticipate that this approach offers a feasible solution for large-scale ground and satellite-based quantum networks~\cite{Gundogan2020}.

\section*{METHODS}

\subsection*{Experimental setup}
The experimental setup (Fig.~\ref{fig:setup}({\bf A})) consists of ATS-memory components (including optical pulse generation and detection systems) and BEC-generation components. As part of the  memory components, probe and control fields are extracted from two independent continuous-wave lasers and then temporally shaped into short pulses using acousto-optic modulators (AOMs). After attenuating the probe beam to the single-photon-level with neutral density filters (NDF) and setting the peak power of the control beam to $8$~mW, both beams are coupled into single-mode optical fibers (FC), and decoupled back to free-space on a separate bench where BEC apparatus is located. Following the polarization control with quarter-wave plates, the probe and control beam diameters are focused to a  waist of $25~\mu$m (or $65~\mu$m) and $150~\mu$m at the intersection of two crossed ODT beams (derived from a 1064 nm laser), using a telescope and single lens respectively (not displayed in the figure).
After coupling into an optical fiber, the output probe pulses are detected using a single-photon detector, and their arrival times are recorded on a time-to-digital converter (TDC), triggered by a function generator.

The ultracold atoms are prepared using standard laser cooling and trapping techniques, similar to Ref.~\cite{Lin2009}. The sequence of these techniques begins with the preparation of cold atoms in a magneto-optical trap (MOT) followed by further sub-Doppler laser cooling to temperatures down to $T\approx50~\mu$K. Next, the atoms are transferred to a quadrupole magnetic trap for RF-induced evaporative cooling that leads to $T<10~\mu$K. Finally, these atoms are transferred into the ODT, which has a controllable trap-depth for further evaporative cooling, and with sufficient cooling, they reach the conditions of Bose-Einstein condensation, as detailed in Supplementary Information. 

\subsection*{Measurements}
In our demonstrations, the memory performance is assessed with time-resolved photon-counting measurements for the detection of both the input-probe photons and stored-and-recalled probe photons (Fig.~\ref{fig:setup}({\bf A})). Each measurement period is performed during a 1-ms-detection window that follows 15-seconds of BEC-preparation and 3.5-ms of free-expansion. In each period, a storage-and-recall event (defined by a pre-set storage time) is repeated $N_{\rm r}=1000$ to 100 times, depending on the storage time $t_{\rm s}$ between 200~ns and 10~$\mu$s, respectively. By repeating these measurement periods several times from $N_{\rm cyc}=10~\rm to~300$, we acquire detection-vs-time histograms for a total number of storage-and-recall events between $N_{\rm A}=N_{\rm r} N_{\rm cyc}=10^4$  and $3\times10^5$, depending on average photon-number of the input probe ($\overline{n}_{\rm in}=0~\rm to~3$).

\subsection*{Memory efficiency and bandwidth}
Here, we detail the current experimental limitations and quantitative predictions for realizing both large memory efficiency and large acceptance bandwidth by exploiting both the favorable efficiency scaling of the ATS protocol and the large optical depth of a BEC. First, we look into the optical depth and bandwidth dependence of ATS-memory~\cite{Saglamyurek2018a}. In the forward-recall configuration (i.e., the input and recalled photons propagate in the same direction, as in our experiments), efficiency is given by 
 \begin{align}
\eta_{\rm f}\approx{(d/2F)}^2 e^{-d/2F} e^{-1/F} \label{eq:effForward}
\end{align}
where $d$ is the peak optical depth, and $F=2\pi B/\Gamma_{\rm eg}$ is the ATS factor that depends on the bandwidth of the probe pulse ($B=0.44/\tau_{\rm p}$ for a Gaussian profile) and the natural linewidth of the optical transition ($\Gamma_{\rm eg}=2\gamma_{\rm eg}$). Equivalently, $F$ relates the duration of the probe pulse $\tau_{\rm p}$ relative to coherence lifetime of the optical transition $1/\gamma_{\rm eg}$, on the condition that the pulse area of the control fields are set to be $2\pi$.   
  
According to this expression, the theoretical maximum efficiency is only $\eta_{\rm f}\approx 40\%$ for the probe bandwidth used in our demonstrations ($B=3.7\Gamma_{\rm eg}/2\pi$), which is, in part, due to the fundamental limitation imposed by the decay time of the ground-to-excited-level coherence (optical decoherence), accounted for by the term $e^{-1/F}$ in Eq.~\ref{eq:effForward}. Although our experimental maximum efficiency ($\eta_{\rm m}=30\%$) is already close to this theoretical maximum, and can even reach it with more optical depth, efficiencies above $40\%$ are only possible with larger bandwidths (shorter pulses), which reduce the impact of optical decoherence ($e^{-1/F}\rightarrow 1$). Further, even with pulses much broader than the transition linewidth ($B\gg\Gamma_{eg}/2\pi$, or equivalently $\tau_{\rm p}\ll1/\gamma_{\rm eg}$, such that $e^{-1/F}\approx 1$), the theoretical efficiency cannot exceed $\eta_{f}\approx54\%$  due to the unavoidable re-absorption of the retrieved probe pulses in the storage medium.
This limitation can be circumvented by implementing ``backward recall'' using counter-propagating write and read-out control pulses~\cite{Gorshkov2007, Saglamyurek2018a}. In this case, the efficiency of ATS memory is  
\begin{align}
\eta_{\rm b}\approx(1-e^{-d/2F})^{2}e^{-1/F}.\label{eq:effBack}
\end{align}
This expression dictates that near-unity memory efficiencies $\eta\geq90\%$ necessitate both sufficiently large bandwidths $B\ge14\Gamma_{\rm eg}/2\pi$ and optical depths $d\ge90$ such that $d/2F\ge3$, which also highlights the inherent broadband character of the ATS protocol. Given that large optical depths are readily available with the BEC platform, we can experimentally achieve such large efficiencies in the backward scheme simply by meeting the following technical requirements: (i) sampling the dense region of BEC with a probe-beam diameter that is smaller than the Thomas-Fermi diameter ($R_{\rm p}\ll R_{\rm TF}$), and (ii) reducing the probe pulse duration, which  also requires using the D1 line in Rb due to the larger hyperfine splitting in the excited state manifold ($\approx 0.8$ GHz). 

Second, to verify this conjecture, we predict the optical depth of our ultracold system based on its experimental characterisations (see Supplementary Information). The effective value of the optical depth depends on the spatial intensity profile of probe as well as the spatial-density profile of the atomic cloud, which is determined by $T$ and ODT parameters. This value can be numerically extracted from Beer's absorption law 
\begin{align}
    d = -\ln\left[\frac{\iint I_{\rm out}(x,y)\: dx dy}{\iint I_{\rm in}(x,y)\: dx dy}  \right], \label{eq:effd}
\end{align}
where $I_{\rm in}(x,y)$ and $I_{\rm out}(x,y)$ are the transverse intensity profiles of the input and transmitted probe, propagating along the $z$-direction. Assuming that the input probe beam is characterized by a Gaussian profile, $I_{\rm in}(x,y)$ and $I_{\rm out}(x,y)$ are then given by
\begin{align}
& I_{\rm in}(x,y) = I_{0}\exp\left(-\frac{2x^{2}}{{R_{p x}}^{2}}\right)\exp\left(-\frac{2y^{2}}{{R_{py}}^{2}}\right) \label{eqn13} \\
& I_{\rm out}(x,y) = I_{\rm in}(x,y)\exp\left[-\frac{3\lambda^{2}}{2\pi}\alpha^{2}\int_{0}^{L} \rho(x,y,z)\: dz,\right] \label{eqn14}
\end{align}
where \{$R_{\rm px}$, $R_{\rm py}$\} are the beam diameters along \{$x,y$\} axes, $I_{0}$ is the peak intensity of the input probe, $\lambda$ is resonant wavelength, $\alpha$ is the strength of the atomic transition, and $\rho(x,y,z)$ is the density distribution of the atomic cloud, involving thermal and BEC components. In this way, we predict $d$ for a given $R_{\rm p}$ as well as $T$ and ODT-trap frequencies, both of which determine the density distributions of BECs and thermal clouds, as further detailed in Supplementary Information.

\subsection*{Memory lifetime}

In this section we detail the limitations of memory lifetime in our demonstrations, and show our predictive calculations for long-lived storage.

In our experiments, storage times are limited to the microsecond timescale due to the decoherence of collective spin excitations. We find that this decoherence is mainly governed by the combined effect of thermal diffusion and magnetic dephasing; the impact of the other decoherence mechanisms (including inelastic collisions~\cite{Dutton2004} and recoil motion~\cite{Dutton2004, Ginsberg2007, Riedl2012}) are predicted to be considerable only at much longer timescales (on the order of milliseconds). The detriment of the thermal diffusion is two-fold: the loss of the atoms due to their dispersive motion out of the interaction cross-section, and the loss of spatial coherence (initially set by the probe and control-wave vectors during writing). While the former is expected to be observable at millisecond and greater time scales, the latter can be observed at much shorter times due to the non-zero probe-control separation angle. In such a configuration, a spatially periodic phase pattern is imprinted as the stored spin-wave with a spatial period of $\lambda{\rm sw}=2\pi/|\boldsymbol{\kappa}_{\rm sw}|$, where $\boldsymbol{\kappa}_{\rm sw} =\mathbf{k}_{\rm p}-\mathbf{k}_{\rm c}$ is imposed by conservation of momentum (phase-matching condition), involving the wavevectors of the probe ($\mathbf{k}_{\rm p}$) and control ($\mathbf{k}_{\rm c}$) beams~\cite{Zhao2008}. Since this phase-grating can be partially or completely erased as a result of the atomic diffusion, thermal-memory lifetime ($\tau_{\rm th}$) depends on $\theta$ (determining the spatial period) and $T$ (determining the diffusion rate), expressed by
\begin{align}
\tau_{\rm th}= \frac {\lambda_{\rm sw}}{2\pi v_{\rm th}}\approx\frac{\lambda}{4\pi\sin{\theta/2}}\sqrt{\frac{m}{k_{\rm B}T}}.\label{eq:dec}
\end{align}
where $v_{\rm th}=\sqrt{{k_{\rm B}T}/{m}}$ is the mean thermal speed, $m$ is the mass of an atom, and $k_{\rm B}$  is the Boltzmann constant. Given that the difference between the wavelength ($\lambda$) of the probe and control fields is very small, $\lambda_{\rm sw}$ is nearly equal to $\lambda/[2\sin(\theta/2)]$,  yielding $|\mathbf{k}_{\rm p}|\approx|\mathbf{k}_{\rm c}|$.

In our experimental configuration with the large probe-control separation angle ($\theta=110^{\circ}$), the impact of thermal diffusion becomes significant at relatively high temperatures ($T>3~\mu$K) and hence dominates over magnetic dephasing in this regime. For example, we expect the thermal decay time constant to be $\tau_{\rm th}=3.2~\mu$s (Eq.~\ref{eq:dec}) at $T=6.2~\mu$K, which is close to the experimentally measured memory lifetime for the same temperature ($4.5~\mu$s). However, in the BEC regime where  $T<0.5~\mu$K, observed memory lifetimes are significantly shorter than those predicted from thermal decoherence, indicating that magnetic dephasing is the dominant decoherence mechanism at ultralow temperatures.  

Based on these characteristics, we describe the observed storage time ($t_{\rm s}$) dependence of memory efficiency for a given cloud-temperature as,
\begin{align}
\eta(t_{\rm s})=&\eta(t_{\rm s0}) e^{-{(t_{\rm s}-t_{\rm s0})}/{\tau_{\rm mag}}} \nonumber \\ &\times\left[\mathcal{F}_{\rm BEC}+\mathcal{F}_{\rm th}e^{-{(t_{\rm s}-t_{\rm s0})^2}/{\tau_{\rm th}^2}}\right], 
\label{eq:decoherence1}
\end{align}
where $\mathcal{F}_{\rm BEC}$ and $\mathcal{F}_{\rm Th}=1-\mathcal{F}_{\rm BEC}$ are the fractions of BEC and thermal atoms in the cloud, respectively, and $\eta(t_{\rm s0})$ is memory efficiency measured for the shortest storage time, $t_{\rm s0}$. In this simplified model, the Gaussian term with the charasteristic decay time of $\tau_{\rm th}$ describes diffusion-induced decoherence (Eq.~\ref{eq:dec}) only for thermal atoms, as the BEC atoms are considered to be free from thermal motion. The common exponential factor with the decay-time constant of  $\tau_{\rm mag}$ corresponds to decoherence due to magnetic dephasing of spin excitations across whole ensemble. We note that in typical cold-atom experiments  with mm-scale ensembles, magnetic dephasing is generally characterized by a Gaussian decay-function (with a time-constant exhibiting an inverse linear dependence on the length of the cloud), instead of the exponential decay term in Eq.~\ref{eq:decoherence1}. We attribute this difference to the fact that in our system, the spatial extent of our atomic ensembles is much smaller and hence exhibits more sensitivity to  variations of ambient magnetic field on micrometer length scales. Moreover, as the atomic cloud falls from the ODT during the 1-ms storage-recall cycle in our single-photon-level measurements, it experiences additional position-dependent magnetic field variations across the  free-fall distance, which is comparable to size of the cloud. 

Using this decoherence model, we fit our results from the measurements of storage time vs. efficiency to Eq.~\ref{eq:decoherence1} (Fig.~\ref{fig:single-photon}({\bf E})), which shows a reasonable agreement with data taken at different cooling temperatures. In the fitting procedure, for a given $T$, we fix the parameters of $\mathcal{F}_{\rm BEC}$, $\mathcal{F}_{\rm th}$~(extracted from independent sets of temperature measurements) and $\tau_{\rm th}$ (calculated from Eq.~\ref{eq:dec} for $\theta=110^{\circ}$) such that $\tau_{\rm mag}$ is a single free parameter that is evaluated from experimental data. We find that  $\tau_{\rm mag}$ tends to be larger with lower cloud-temperatures because of the fact that the spatial extension of atomic clouds become  smaller and thus less susceptible to the variations of ambient magnetic fields. 

We also note that the extracted value of $\tau_{\rm mag}$ exhibits a variation on long timescales (over several days), which we attribute to changes in the conditions of ambient magnetic-field. For instance, we find that $\tau_{\rm mag}$ varies between $5$ and  $7~\mu$s for a cloud at $T=340$~nK, depending on day-to-day optimization of bias magnetic field.  Figure.~\ref{fig:single-photon}({\bf E}) represents one of the best data sets, obtained after a careful bias-field optimization, yielding magnetic dephasing constant of $\tau_{\rm mag} =$ ($7.0 \pm 2.5$, $7.0 \pm 1.0$, $16.5 \pm 2.8$) $ \mu$s  with $1/e$ memory-lifetime of $\tau_{\rm m} =$ ($4.5 \pm 2.5$, $7.8 \pm 1.0$, $15.8 \pm 2.8$) $\mu$s for $T = (6200$, $340$, $280$) nK, respectively. In addition, we occasionally observe that memory efficiency drifts by about $10-30~\%$ of the typical $\eta_{\rm m}$ (in part due to instability of the atom-number of prepared clouds in hours-long time scales), which seems to be more pronounced in measurements for short storage times ($\tau_{\rm s}<2~\mu$s) carried out for a BEC cloud.

 Finally, we show the details of our predictions for the ultimately achievable memory lifetimes ($\tau_{\rm m}$) in our current system. These predictions are based on our experimentally realised clouds (characterised by $T$, $\mathcal{F}_{\rm BEC}$ and $\mathcal{F}_{\rm th}$) at a  probe-control separation angle $\theta$,  under the assumption that the decoherence effects such as magnetic dephasing and BEC-spatial decoherence can be eliminated by technical means. In this case, we consider the impact of three major decoherence mechanisms: (i) thermal motion, (ii) recoil motion, and (iii) inelastic two-body collisions. The effects of thermal- and recoil-motion induced decoherence are separately described by Gaussian decays of memory efficiency with characteristic times of $\tau_{\rm th}$ (Eq.~\ref{eq:dec}) and $\tau_{\rm rec}=R_{\rm p} \lambda  m / [2h  \sin(\theta/2)]$~\cite{Lettner2011, Riedl2012}, respectively. While thermal-motion induced decoherence is not applicable to the BEC part of atomic cloud, the effect of recoil-motion is ignored for the thermal part (as being significantly dominated by thermal decoherence) in our regime of interest. Furthermore, decoherence due to inelastic two-body collisions in BEC is characterised by an exponential decay of memory efficiency with a decay time of $\tau_{\rm col}=m/[4h I_{\rm m} (a_{\rm sc}) \rho_{\rm B}]$, where $I_m (a_{\rm sc})$ is the imaginary part of scattering length for the two component Rb-BEC (in $\ket{\rm g}$ and $\ket{\rm s}$), and $\rho_{\rm B}$ is the peak density of BEC~\cite{Dutton2004,Zhang2009}. As the atomic density in a thermal cloud is substantially smaller than a BEC, $\tau_{\rm col}$ is considered to be negligibly small for the thermal portion of the cloud. 

Under these conditions, the combined effect of the decoherence mechanisms on memory efficiency $\eta$ is described by 
\begin{align}
\eta(t_{\rm s})=\eta(0) \Bigg[&\mathcal{F}_{\rm BEC}\left(e^{-t_{\rm s}/ \tau_{\rm col}}\right) \left(e^{-{t_{\rm s}}^2/{\tau_{\rm rec}}^2}\right)\nonumber\\
&+(1-\mathcal{F}_{\rm BEC})\left(e^{-{t_{\rm s}}^2/{\tau_{\rm th}}^2}\right)\Bigg], 
\label{eq:decoher}
\end{align}
where $t_{\rm s}$ is storage time, and $\eta(0)$ is memory efficiency for $t_{\rm s0}=0$. The memory lifetime shown in Fig.~\ref{fig:performance}({\bf C}) is defined as the characteristic time at which the memory efficiency drops to ($1/e$) of $\eta(0)$.

\subsection*{Four-wave mixing noise}
FWM noise is one of the main limitations towards developing reliable broadband spin-wave memories. The probability of FWM noise depends on two general factors: (i) geometric relationships between the wavevectors of probe fields and control fields, and (ii) memory resources including optical depth and control-field power along with system specific parameters. 

The geometric factor is characterised by the angular separation $\theta$ between the probe and control fields. FWM noise is detrimental to memory for a $\theta \le \theta_{\rm FWM}$, which that satisfies the general phase-matching conditions, as detailed in Supplementary Information. This threshold angle is $\theta_{\rm FWM}=2\arcsin(\sqrt{{\lambda}/[{8\pi L}]})$, for an effective medium length of $L$ along the propagation direction of the probe field. By choosing $\theta\geq\theta_{\rm FWM}$, FWM noise can be mostly eliminated, at the expense of larger thermal-diffusion-induced decoherence. 

The resource dependence of FWM is characterised by a ``noise strength'' parameter, which is proportional to the probability of  FWM noise corrupting the memory, as detailed in Ref.~\cite{Lauk2013, Geng2014a}. The FWM noise-strength is determined by optical depth $d$, peak Rabi frequency of the control field $\Omega_{\rm c}$, and system-specific parameters (e.g.~$\gamma_{ge}$ in a $\Lambda$-type three-level system) as in the following,
\begin{equation}
S_{\rm FWM}\propto \Omega_{\rm c}^4\left[\sinh\left(\frac{\zeta d \gamma_{\rm eg}}{\Delta_{\rm gs}}\right)\right]^2, \label{eq:FWM}
\end{equation}
where $h\Delta_{\rm gs}$ is the energy difference between the ground levels of the $\Lambda$-system, and $\zeta={\Omega_{\rm c}}/{\Omega_{\rm c}^{\prime}}$ is the ratio of the Rabi frequency from $\ket{s}\rightarrow\ket{e}$ to the one from $\ket{g}\rightarrow\ket{e}$. Since the required  $d$ and $\Omega_{\rm c}$ for implementing an optimal broadband memory are proportional to the memory bandwidth ($B>\Gamma_{\rm eg}/2\pi$) with certain proportionality constants specific to the memory protocol, FWM noise strength strongly depends on bandwidth and the employed protocol. The ATS protocol is advantageous for eliminating  FWM noise due to its favorable resource scaling in the broadband operation regime, as compared to adiabatic memories, such as EIT and off-resonant Raman protocols. In Fig.~\ref{fig:performance}({\bf D}), we make a comparison of FWM noise-strength ($S_{\rm FWM}$) between the ATS and EIT protocols for implementation of an optimal broadband memory in our $^{87}$Rb system (featuring $\Delta_{\rm gs}/2\pi=$6.83 GHz and $\zeta\approx1.33$). In this comparison, we calculate $S_{\rm FWHM}$ using Eq.~\ref{eq:FWM} for a bandwidth range of $10(\Gamma_{\rm eg}/2\pi)<B<40(\Gamma_{\rm eg}/2\pi)$, requiring optical depths of $d_{\rm ATS}=8\times(2\pi B/\Gamma_{\rm eg})$ and $d_{\rm EIT}=50\times(2\pi B/\Gamma_{\rm eg})$ as well as peak Rabi frequencies of $\Omega_{\rm ATS}=1.5\times (2\pi B)$ and $\Omega_{\rm EIT}=4\times (2\pi B)$ for optimal ATS and EIT memories, based on their non-adiabatic and adiabatic operation conditions, respectively (see Ref.~\cite{Rastogi2019} for details). These results show that in this bandwidth range, corresponding to probe pulse durations between $1.9-7.3$~ns in our Rb system, the probability of FWM noise associated with an optimal ATS memory is 4--5 orders of magnitude smaller than that associated with an optimal EIT memory. Consequently, the ATS protocol offers a favorable option for the realization of long-lived broadband quantum memories featuring both high-speed and faithful operation. 

\subsection*{Acknowledgments}

We thank Dr. Khabat Heshami for useful discussions, and appreciate generous technical support from Paul Davis and Greg Popowich.  We gratefully acknowledge funding from the Natural Science and Engineering Research Council of Canada (NSERC RGPIN-2014-06618), Canada Foundation for Innovation (CFI), Canada Research Chairs Program (CRC), the Alberta Major Innovation Fund Quantum Technologies project, Alberta Innovates, and the University of Alberta.


\begin{thebibliography}{10}

\bibitem{Lvovsky2009b}
A.~I. Lvovsky, B.~C. Sanders, and W. Tittel, Nat. Photonics {\bf 3},  706
  (2010).

\bibitem{Heshami2016c}
K. Heshami {\it et~al.}, J. Mod. Opt. {\bf 63},  2005  (2016).

\bibitem{Phillips2001b}
D.~F. Phillips {\it et~al.}, Phys. Rev. Lett. {\bf 86},  783  (2001).

\bibitem{Liu2001a}
C. Liu {\it et~al.}, Nature {\bf 409},  490  (2001).

\bibitem{Turukhin2002a}
A.~V. Turukhin {\it et~al.}, Phys. Rev. Lett. {\bf 88},  023602  (2001).

\bibitem{Boozer2007}
A.~D. Boozer {\it et~al.}, Phys. Rev. Lett. {\bf 98},  193601  (2007).

\bibitem{Wilk2007}
T. Wilk {\it et~al.}, Science (80-. ). {\bf 317},  488  (2007).

\bibitem{Fleischhauer2000b}
M. Fleischhauer and M.~D. Lukin, Phys. Rev. Lett. {\bf 84},  5094  (2000).

\bibitem{Nunn2007}
J. Nunn {\it et~al.}, Phys. Rev. A {\bf 75},  011401  (2007).

\bibitem{Gorshkov2007}
A.~V. Gorshkov {\it et~al.}, Phys. Rev. A {\bf 76},  033805  (2007).

\bibitem{Moiseev2001}
S.~A. Moiseev and S. Kr{\"{o}}ll,   17  (2001).

\bibitem{Afzelius2009a}
M. Afzelius {\it et~al.}, Phys. Rev. A {\bf 79},  052329  (2009).

\bibitem{Dudin2013}
Y.~O. Dudin, L. Li, and A. Kuzmich, Phys. Rev. A {\bf 87},  31801  (2013).

\bibitem{Heinze2013}
G. Heinze, C. Hubrich, and T. Halfmann, Phys. Rev. Lett. {\bf 111},  033601
  (2013).

\bibitem{Hedges2010}
M.~P. Hedges {\it et~al.}, Nature {\bf 465},  1052  (2010).

\bibitem{Hosseini2011c}
M. Hosseini {\it et~al.}, Nat. Commun. {\bf 2},  174  (2011).

\bibitem{Hsiao2018a}
Y.-F. Hsiao {\it et~al.}, Phys. Rev. Lett. {\bf 120},  183602  (2018).

\bibitem{Reim2010b}
K.~F. Reim {\it et~al.}, Nat. Photonics {\bf 4},  218  (2010).

\bibitem{Guo2018}
J. Guo {\it et~al.}, Nat. Commun. {\bf 10},  148  (2019).

\bibitem{Gundogan2015}
M. G{\"{u}}ndoƒüan {\it et~al.}, Phys. Rev. Lett. {\bf 114},  230501  (2015).

\bibitem{Ding2015a}
D.-S. Ding {\it et~al.}, Nat. Photonics {\bf 9},  332  (2015).

\bibitem{Vernaz-Gris2018}
P. Vernaz-Gris {\it et~al.}, Nat. Commun. {\bf 9},  363  (2018).

\bibitem{Wang2019b}
Y. Wang {\it et~al.}, Nat. Photonics {\bf 13},  346  (2019).

\bibitem{Saglamyurek2018a}
E. Saglamyurek {\it et~al.}, Nat. Photonics {\bf 12},  774  (2018).

\bibitem{VestergaardHau1999a}
L.~V. Hau {\it et~al.}, Nature {\bf 397},  594  (1999).

\bibitem{Dutton2004}
Z. Dutton and L.~V. Hau, Phys. Rev. A {\bf 70},  053831  (2004).

\bibitem{Zhang2009}
R. Zhang, S.~R. Garner, and L.~V. Hau, Phys. Rev. Lett. {\bf 103},  233602
  (2009).

\bibitem{Ginsberg2007}
N.~S. Ginsberg, S.~R. Garner, and L.~V. Hau, Nature {\bf 445},  623  (2007).

\bibitem{Lettner2011}
M. Lettner {\it et~al.}, Phys. Rev. Lett. {\bf 106},  210503  (2011).

\bibitem{Riedl2012}
S. Riedl {\it et~al.}, Phys. Rev. A {\bf 85},  022318  (2012).

\bibitem{Rastogi2019}
A. Rastogi {\it et~al.}, Phys. Rev. A {\bf 100},  012314  (2019).

\bibitem{Lauk2013}
N. Lauk, C. O'Brien, and M. Fleischhauer, Phys. Rev. A {\bf 88},  013823
  (2013).

\bibitem{Geng2014a}
J. Geng {\it et~al.}, New J. Phys. {\bf 16},  113053  (2014).

\bibitem{Saglamyurek2019c}
E. Saglamyurek {\it et~al.}, Phys. Rev. Res. {\bf 1},  022004  (2019).

\bibitem{Lin2009}
Y.-J. Lin {\it et~al.}, Phys. Rev. A {\bf 79},  063631  (2009).

\bibitem{Jenkins2006}
S.~D. Jenkins {\it et~al.}, Phys. Rev. A {\bf 73},  021803  (2006).

\bibitem{Matsukevich2006}
D.~N. Matsukevich {\it et~al.}, Phys. Rev. Lett. {\bf 96},  033601  (2006).

\bibitem{Wang2011}
H. Wang {\it et~al.}, Phys. Rev. A - At. Mol. Opt. Phys.  (2011).

\bibitem{Farrera2018}
P. Farrera, G. Heinze, and H. de~Riedmatten, Phys. Rev. Lett. {\bf 120},
  100501  (2018).

\bibitem{Zhao2008}
B. Zhao {\it et~al.}, Nat. Phys. {\bf 5},  95  (2008).

\bibitem{Elliott2018}
E.~R. Elliott {\it et~al.}, npj Microgravity {\bf 4},  16  (2018).

\bibitem{Gundogan2020}
M. G{\"{u}}ndogan {\it et~al.}, arxiv.org  2006.10636  (2020).

\end{thebibliography}

\pagebreak
\pagebreak
\newpage
\begin{widetext}

\begin{center}
{\large \bf Supplementary information to ``Storing short pulses of single-photon-level light in Bose-Einstein condensate for high-performance quantum memory''}

\end{center}
\end{widetext}

\setcounter{equation}{0}
\setcounter{figure}{0}
\setcounter{table}{0}
\setcounter{page}{1}
\makeatletter
\renewcommand{\theequation}{S\arabic{equation}}
\renewcommand{\thefigure}{S\arabic{figure}}
\renewcommand{\thetable}{S\arabic{table}}
\renewcommand{\thesection}{S\Roman{section}}
\renewcommand{\thesubsection}{s\roman{subsection}}

\section{BEC preparation and characterisation}

Our experiments are performed with an apparatus designed for  Bose-Einstein condensate (BEC) production using $^{87}$Rb. 
From a diffusive oven, we pre-cool atoms in a 2D magneto-optical trap (MOT) before guiding them (using a push beam) through 15~cm of differential pumping to the ``science'' chamber.  Here, the atoms undergo further 15~s of laser cooling in a standard 3D MOT, and the general procedure largely follows that of Ref.~\cite{Lin2009}.   In short, a six-beam MOT collects and cools the atoms, followed by a 20-ms near-resonant sub-Doppler optical molasses phase.  Using a short pulse of ``repump'' light on the $\ket{F = 1}\rightarrow \ket{F^\prime = 2}$ transition, atoms are optically pumped into the $F=1$ ground state and magnetically trapped using the quadrupole field of a pair of anti-Helmholtz coils.  Radio-frequency evaporative cooling is performed here until the atoms reach a few microkelvin, at which point an optical dipole trap (ODT) derived from a pair of 90-degree-intersecting, 1064~nm focussed laser beams is turned on.  By positioning this trap just below the zero of the magnetic quadrupole field, atoms are prevented from experiencing significant Majorana losses, and the cloud is efficiently transferred to the ODT.  In this trap, forced evaporative cooling further proceeds by lowering the power in the ODT beams to 11\% of its initial value, producing the coldest clouds in our experiments.

\begin{figure*}[tb]
\begin{center}
\includegraphics{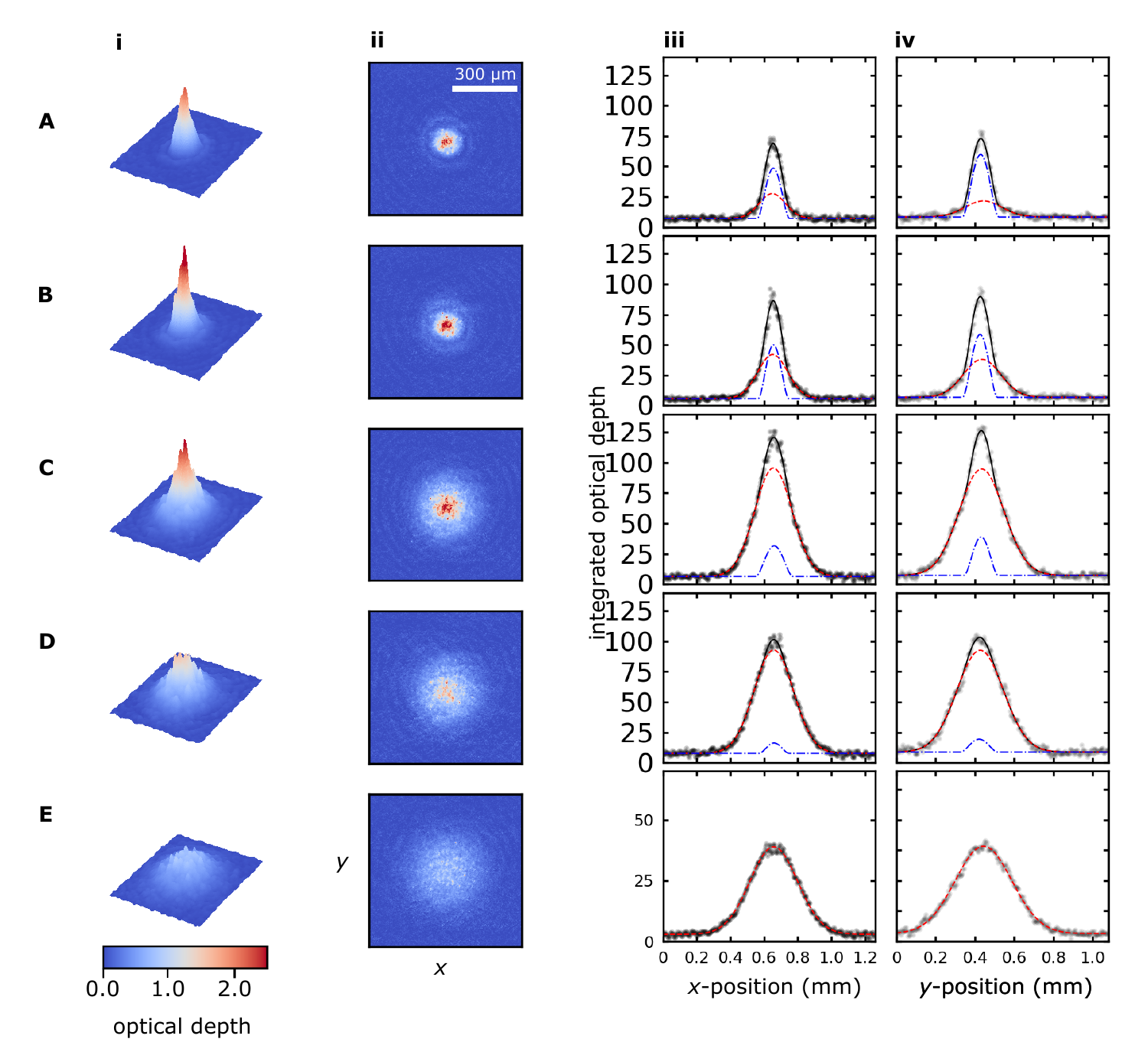}
\caption{\textbf{BEC characterization, representative data.} Images and fits for 20~ms time-of-flight absorption profiles of atomic samples after various evaporation stop-points.  Columns show: ({\bf i}) 3D rendering of absorption image, as in Fig.~\ref{fig:setup}({\bf B}) of the main text; ({\bf ii}) Absorption map of atomic density, calculated in dimensional optical density units (scale bar is common to all and shown in ({\bf Aii})); ({\bf iii}) Optical density integrated (summed) along the $y$-axis to give profile along $x$, plus bimodal fits showing fully bimodal fit (black) and individual components of this fit, including a Gaussian (red dashed) and Thomas-Fermi profile (blue dash-dot); ({\bf iv}) Optical density integrated (summed) along the $x$-axis to give profile along $y$ with the same fits as ({\bf iii}).  The rows show different temperature clouds from top to bottom, with the temperatures extracted from the Gaussian fits.  These representative images are part of 10-image data sets used for calibration, which gave temperatures: ({\bf A}) $280 \pm 50 ~\mu$K; ({\bf B}) $240 \pm 15 ~\mu$K; ({\bf C}) $340 \pm 5 ~\mu$K; ({\bf D}) $370 \pm 8 ~\mu$K; ({\bf E}) $505 \pm 10 ~\mu$K. The uncertainties are the calculated standard deviation from the ten fits. The temperature for ({\bf A}) may be suspect due to the small Gaussian component used to determine this temperature.}
\label{fig:BECdata}
\end{center}
\end{figure*}

To characterise the temperature dependence of the memory storage processes, we cool the atoms to different ODT evaporation end-points.  We calibrated these temperatures and atom numbers by performing time-of-flight measurements on these samples: we release the atoms from their ODT and capture absorption images after 20~ms TOF.  The atomic distributions are measured by dividing an image with atoms present by the one with the absorbing beam alone, followed by taking the negative natural logarithm to calculate the optical density distribution (Fig.~\ref{fig:BECdata}({\bf ii})).  These optical density images are then summed numerically along either the $x$- or $y$-axis, and the resulting 1D distributions are fit to the appropriate models.  In the case of only thermal atoms, we fit a Gaussian distribution to the data, and use the width to extract temperature and atom number. For cases where both thermal and BEC atoms are present, we use a two-Gaussian fit to best determine the atom number in each component, and a bimodal Thomas-Fermi plus Gaussian fit to get an accurate overall fit, then using the Gaussian part of the bimodal fit to extract the temperature of the thermal component (Figs.~\ref{fig:BECdata}({\bf iii}) and \ref{fig:BECdata}({\bf iv})).

\begin{figure}[tb]
\begin{center}
\includegraphics{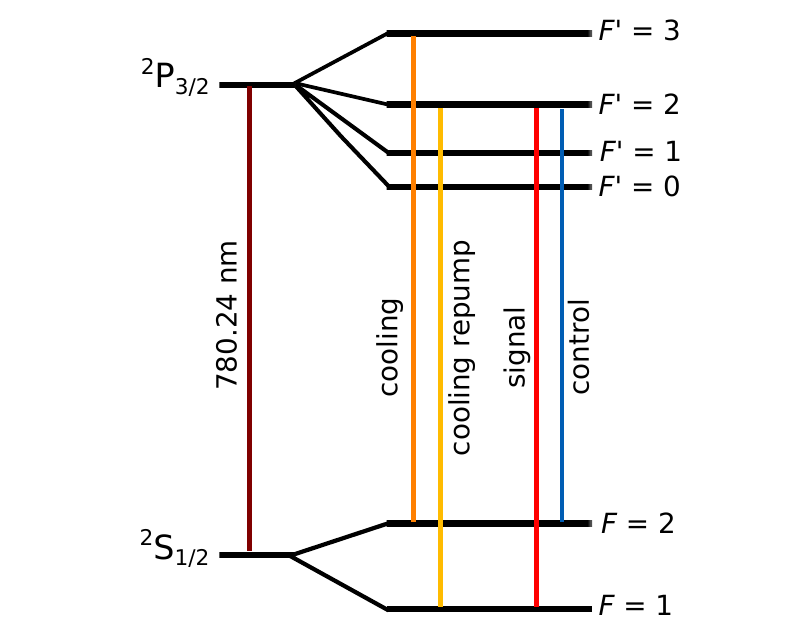}
\caption{\textbf{Rubidium 87 level structure.} Relevant laser transitions labelled, left-to-right: Bare D2 transition is 780.24 nm; cooling transition used for MOT; repump cooling transition used for MOT; signal transition $\ket{F = 1}\rightarrow \ket{F^\prime = 2}$; control transition $\ket{F = 2}\rightarrow \ket{F^\prime = 2}$}
\label{fig:rblines}
\end{center}
\end{figure}

\section{Estimating the threshold separation angle for four-wave-mixing noise} \label{FWM angle}

Four wave-mixing (FWM) noise is one of the fundamental limitations for the reliable operation of a broadband spin-wave quantum memory. It is possible to suppress this noise partially or completely by a proper choice of the separation angle between probe and control ($\theta$). In this section, we derive an estimate the threshold value of $\theta$, above which  FWM noise is significantly suppressed.

\begin{figure*}
\begin{center}
\includegraphics[width = 170 mm] {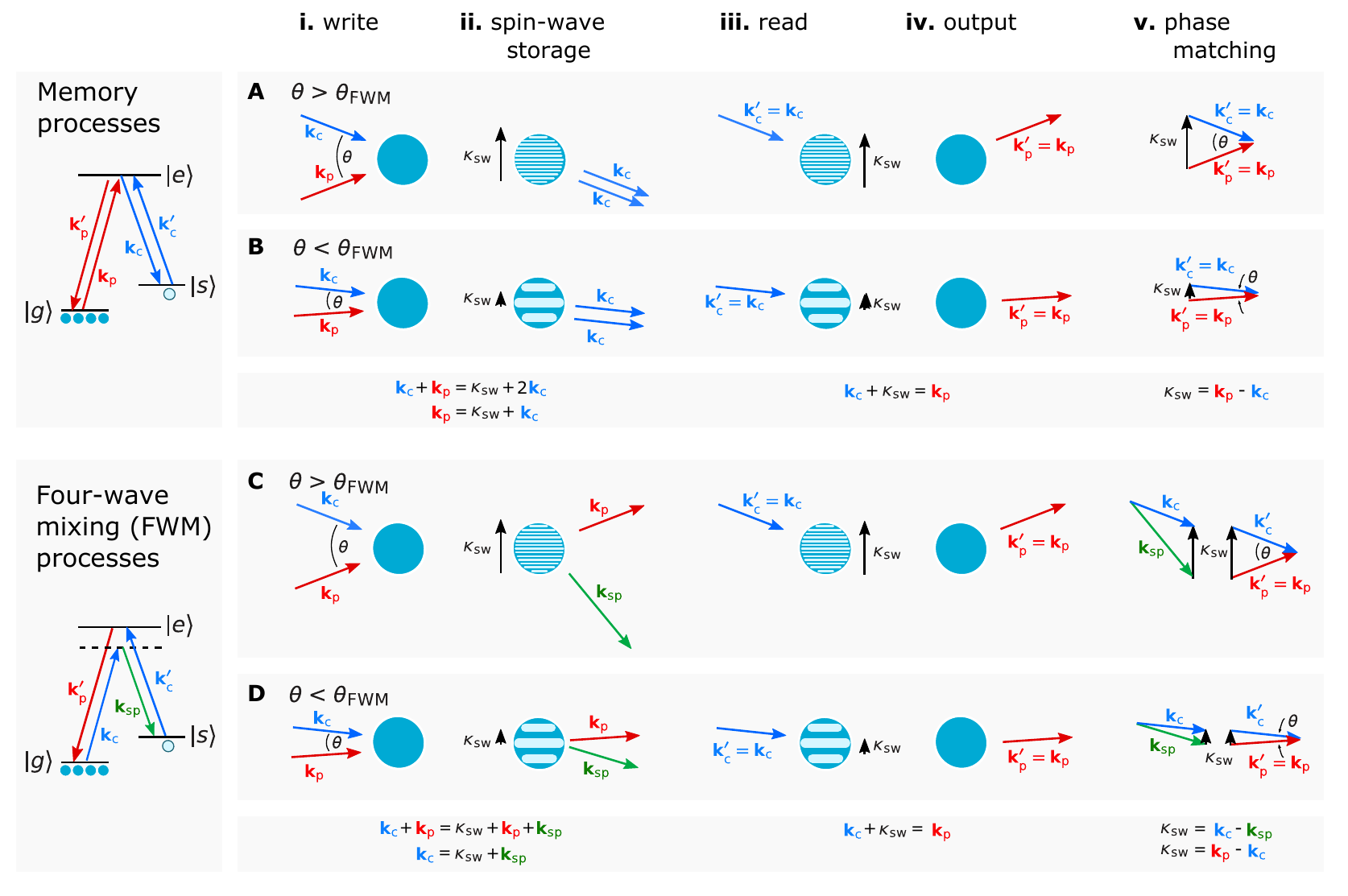}
\caption{
\textbf{Phase matching and momentum conservation in memory ({\bf A}, {\bf B}) and four-wave mixing ({\bf C}, {\bf D}) processes.}  Left panels: Energy level diagrams showing the transitions from $\ket{g}$ to $\ket{s}$ involved in write processes and from $\ket{s}$ to $\ket{g}$ in read processes, for memory (upper) and FWM (lower).  In both cases, the energy-matching condition is satisfied for the two two-photon transitions involved in the write and read processes. For large ({\bf A,C}) and small ({\bf B,D}) angular separations of the input probe and control fields, Column ({\bf i}) shows the write process, ({\bf ii}) shows the storage process (including the output signals from ({\bf i}), with cartoon spin-wave representations shown as stripes with different spacings, {\bf iii}) shows the read process, and ({\bf iv}) shows the output.  Column ({\bf v}) shows the wavevector additions for the write-to-storage and read-to-output processes.  For the memory processes {\bf A}, {\bf B}, the probe photon is absorbed in the write stage ({\bf i}), and a photon is stimulated into the control mode to complete the two-photon transition ({\bf ii}).  When the same control mode is used for reading ({\bf iii}), the output probe photon ({\bf iv}) is emitted into the probe mode, and the momentum-matching condition for the write and read processes ({\bf v}) are identical.  In considering noise, we look only at FWM processes ({\bf C}, {\bf D}) that result in the \emph{same} spin-waves as the desired memory processes, and thus, the identical read/output processes ({\bf iii,iv}). In these cases, the probe photon is not absorbed, and instead, a control photon off-resonantly drives the first transition.  A spontaneous photon completes the two-photon transition, to create the spin-wave $\boldsymbol{\kappa}_{\rm sw}$ identical to the one from the memory process.  Depending on the input angle $\theta$, the direction and magnitude of this spontaneous photon in mode $\mathbf{k}_{\rm sp}$ must ``adjust'' to meet the momentum-matching condition ({\bf v}, left); when this vector is stretched to meet this condition, the frequency of the mode is necessarily increased, which may result in a violation of the energy-matching condition.  Thus, for large angles, the phase-matching condition is not satisfied and FWM noise is suppressed.
}  
\label{fig:spinwaves}
\end{center}
\end{figure*}

A generic spin-wave memory for a three-level atom-like system (left side, Fig.~\ref{fig:spinwaves}) is operated (regardless of the  protocol employed) by applying a pair of electromagnetic fields at the ``write'' stage, relying on the formation of a spin-wave in the atomic coherence, and reapplying the control field at the ``read'' stage to observe an output signal within the mode of the initial probe field.  Here, we consider the ``mode'' to be the spatial and energetic modes of the field.  In considering FWM noise, we consider both the desired memory process (Fig.~\ref{fig:spinwaves}({\bf A,B})) and the undesired noise process (Fig.~\ref{fig:spinwaves}({\bf C,D})) that lead to outputs involving a signal in the probe mode upon performance of the generic protocol, and we consider how the angle between the two input modes affects the probability of the undesired FWM process producing a field in the signal/probe mode.

In processes leading to memory, the probe field, characterised by wavevector $\mathbf{k_{\rm p}}$  (Fig.~\ref{fig:spinwaves}{\bf{Ai,Bi}}), drives the $\ket{g}$ to $\ket{e}$ transition while the control field, characterised by wavevector $\mathbf{k_{\rm c}}$, drives the $\ket{e}$ to $\ket{s}$ transition (upper left panel in Fig.~\ref{fig:spinwaves}), and a two-photon transition creates an atomic coherence between ground states with a spatial periodicity characterised by a spinwave wavevector $\boldsymbol{\kappa}_{\rm sw}$.  The two-photon process that creates the spin-wave can be described as the absorption of the probe field and stimulated emission into the control field, such that the conservation of momentum from the read to the storage stages reads
\begin{align}
    \mathbf{k}_{\rm p} + \mathbf{k}_{\rm c} &= \boldsymbol{\kappa}_{\rm sw} + 2\mathbf{k}_{\rm c} \nonumber \\
    \mathbf{k}_{\rm p} &= \boldsymbol{\kappa}_{\rm sw} + \mathbf{k}_{\rm c}.
\end{align}
The wavelength associated with the spinwave wavevector $\boldsymbol{\kappa}_{\rm sw}$, $\lambda_{\rm sw} = 2\pi/|\boldsymbol{\kappa}_{\rm sw}|$ depends on the angle between the incoming probe and control fields, as depicted schematically in Fig.~\ref{fig:spinwaves}{\bf{Aii,Bii}}: a small angle $\theta$ creates long-wavelength spin wave and short wavevector.

In contrast, the FWM noise process can create the same spin-wave (characterised by the same $\boldsymbol{\kappa}_{\rm sw}$ by an alternate process that involves spontaneous emission rather than stimulated emission, often referred to a ``Stokes'' process/photon.  In this process, the control field off-resonantly drives the $\ket{g}$ to $\ket{e}$ transition and a spontaneous photon with wavevector $\mathbf{k}_{\rm sp}$ connects $\ket{e}$ to $\ket{s}$ and completes the transfer of atomic coherence (lower left panel in Fig.~\ref{fig:spinwaves}), setting up the ground-state coherence and the spin-wave.  In this case, the probe field does not interact with the system, and the conservation of momentum from the read to the storage stage dictates that  
\begin{align}
    \mathbf{k}_{\rm p} + \mathbf{k}_{\rm c} &= \boldsymbol{\kappa}_{\rm sw} + \mathbf{k}_{\rm p} + \mathbf{k}_{\rm sp} \nonumber \\
    \mathbf{k}_{\rm c} &= \boldsymbol{\kappa}_{\rm sw} + \mathbf{k}_{\rm sp} \label{eq:FWMphasematch}.
\end{align}
Note that though the spontaneous photon may emit into any spatial mode, the only case where it is an issue for noise is the case where it satisfies the above condition and creates the same spin wave as the desired memory process.

For the read stage (Fig.~\ref{fig:spinwaves}{\bf ({\bf iii})}) in the forward-recall configuration (as implemented in this and several past experiments), a control field $\mathbf{k}^\prime_{\rm c}$ identical to the one used in the read stage  is applied to the ensemble ($\mathbf{k}^\prime_{\rm c} = \mathbf{k}_{\rm c}$), and in all cases, this results in the output of a signal (Fig.~\ref{fig:spinwaves}({\bf iv})) into the probe mode $\mathbf{k}_{\rm p}^\prime$ that is identical to the input probe mode: $\mathbf{k}_{\rm p}^\prime = \mathbf{k}_{\rm p}$. The conservation of momentum from the read to the output process is the same for all cases:
\begin{align}
    \mathbf{k}_{\rm c}^\prime + \boldsymbol{\kappa}_{\rm sw} &= \mathbf{k}_{\rm p}^\prime \nonumber \\
    \mathbf{k}_{\rm c}+ \boldsymbol{\kappa}_{\rm sw} &= \mathbf{k}_{\rm p} \label{eq:phasematch}
\end{align}
We note the results that follow apply equally to the backward-recall scheme ($\mathbf{k}_{\rm c}^\prime=-\mathbf{k}_{\rm c}$ and $\mathbf{k}_{\rm p}^\prime=-\mathbf{k}_{\rm p}$) due to symmetry considerations.
For the FWM case, had the spontaneous photon been emitted into a different mode that the one considered here, the output probe photon, often referred to as the ``anti-Stokes'' photon, would be emitted into a different output mode, would not be detected by a detector set up to find a signal in $\mathbf{k}_{\rm p}$, and thus would not contribute to noise of the desired signal. The FWM-noise photons we do consider cannot be discriminated from the output probe photons with standard spectral and temporal filtering techniques, thereby severely hampering the reliable memory operation in the quantum regime. 

For the memory processes, the phase-matching condition is readily satisfied for both the write and read stages, for all choices of $\theta$.  With the same modes used for writing and reading, Eq.~\ref{eq:phasematch} applies in both cases, as seen in the vector addition diagram in (Fig.~\ref{fig:spinwaves}({\bf Av,Bv})).  For FWM processes, phase-matching is an important consideration: in order to establish the spin-wave wavevector that results in a reading process that mimics the memory process and is described by Eq.~\ref{eq:phasematch}, the writing process requires the condition described by Eq.~\ref{eq:FWMphasematch}. Through the equality of $\boldsymbol{\kappa}_{\rm sw}$, this means that the FWM process must satisfy
\begin{align}
     \mathbf{k}_{\rm c} - \mathbf{k}_{\rm sp} &=\mathbf{k}_{\rm p} - \mathbf{k}_{\rm c} \nonumber\\
     \implies 2\mathbf{k}_{\rm c}  &=\mathbf{k}_{\rm p} + \mathbf{k}_{\rm sp}.
\end{align}
As can be seen through the vector addition diagrams in (Fig.~\ref{fig:spinwaves}({\bf Cv,Dv})), when the probe and control modes are of a well-defined frequency/wavelength, the magnitude of the spontaneous mode wavevector $\mathbf{k}_{\rm sp}$ must adjust to satisfy momentum conservation.  For large angular separation between the probe and control beams,  $|\mathbf{k}_{\rm sp}|$ becomes large, implying that the frequency of this mode is smaller.  However, to satisfy the energy-matching component of the phase-matching condition, 
\begin{align}
    \hbar |\mathbf{k}_{\rm c}| - \hbar |\mathbf{k}_{\rm sp}| + \hbar |\mathbf{k}_{\rm c}| - \hbar |\mathbf{k}_{\rm p}| = 0 \label{eq:energymatch}
\end{align}
there is no freedom to vary $|\mathbf{k}_{\rm sp}|$, and so this FWM process is, overall, suppressed for large angles.

To determine the range of $\theta$ values that lead to a significant FWM signal in the memory-output channel, we consider a condition for phase-matching under which any phase mismatch happens on a length scale larger than the effective length $L$ of the storage medium, such that
\begin{align}
|\Delta\mathbf{k}_{\rm FWM}|L\ll1, \label{eqn9}
\end{align}
where 
\begin{align}
\Delta\mathbf{k}_{\rm FWM} =&   |\mathbf{k}_{\rm c}| - |\mathbf{k}_{\rm sp}| + |\mathbf{k}_{\rm c}|^\prime - |\mathbf{k}_{\rm p}| \nonumber\\
 & =  2|\mathbf{k}_{\rm c}| - (|\mathbf{k}_{\rm sp}| +  |\mathbf{k}_{\rm p}|)
\end{align}
is basically the difference of the energy-matching condition (Eq.~\ref{eq:energymatch}) from zero.
Since the optical transitions are much higher energy than the ground state splitting, we assume that each wavevector has approximately the same magnitude  $2\pi/\lambda$, where $\lambda$ is the wavelength of the optical transition. In this case,  $|\Delta\mathbf{k}_{\rm FWM}|=2(2\pi/\lambda)(1 - \cos\theta)$ and the condition for efficient FWM  is  
\begin{align}
& \left(\frac{4\pi L}{\lambda}\right)\left(1 - \cos\theta\right ) \ll 1 , \nonumber \\
& \frac{8\pi L}{\lambda}\sin^{2}(\theta/2)  \ll 1, \nonumber \\
& \theta \ll 2\arcsin\sqrt{\frac{\lambda}{8\pi L}}.  \label{eqn11}
\end{align}

This inequality shows that FWM noise is most detrimental to memory  when the angular separation $\theta$ between the probe and control fields is  much smaller than the FWM threshold angle $\theta_{\rm FWM}=2\sin^{-1}\sqrt{{\lambda}/{8\pi L}}$. For angles $\theta > \theta_{\rm FWM}$, FWM optical fields are not phase-matched and hence the noise is significantly suppressed. Finally, one must consider that although a large angular probe-control separation is favourable for reliable memory operation, it also induces additional spin-wave decoherence mechanisms, as discussed in the main text. 

\section{Calculations for effective optical depth }
The fraction of input probe power that is absorbed as the beam propagates through an atomic cloud can be calculated using Beer's  law (assuming the probe intensity is below the atom's saturation intensity) to give an overall optical density

\begin{align}
    d = \ln\left[\frac{\iint I_{\rm out}(x,y)\: dx dy}{\iint I_{\rm in}(x,y)\: dx dy}  \right], \label{eqn12}
\end{align}
where $I_{\rm in}(x,y)$ and $I_{\rm out}(x,y)$ are the transverse intensity profiles of the input and transmitted probe, propagating along the $z$-direction, and the integrations are assumed to extend over the entire distribution.  The measurements effectively perform the integration by measuring the total power, which is proportional to intensity.  For the Gaussian beams, as in our experiments, the intensity profiles are
\begin{align}
& I_{\rm in}(x,y) = I_{0}\exp\left(-\frac{x^{2}}{2{w_{ x}}^{2}}\right)\exp\left(-\frac{y^{2}}{2{w_{ y}}^{2}}\right) \label{eqn13a} \\
& I_{\rm out}(x,y) = I_{\rm in}(x,y)\exp\left[-d_0(x,y)\right] \label{eqn14a}
\end{align}
where \{$2 w_{\rm x}$, $2 w_{\rm y}$\} are the beam waists  along \{$x,y$\} axes, $I_{0}$ is the peak intensity, and $d$ is the optical depth.  For resonant light, the optical depth  depends on the the absorption wavelength $(\lambda)$, the strength of the atomic transition  ($\alpha$), and the line-integrated atomic density distribution $\rho^{\rm 2D}(x,y) = \int_{0}^{L} \rho(x,y,z)\: dz$. Thus, for a $J \rightarrow J'$ transition of an alkali atom we have:
\begin{align}
& d_{\rm 0} = \frac{3\lambda^{2}}{2\pi}\alpha^{2}\left(\frac{2J' + 1}{2J + 1}\right)\int_{0}^{L} \rho(x,y,z)\: dz, \label{eqn15}     
\end{align}
where $(2J' + 1)/(2J + 1)$ is the degeneracy factor and $\alpha$ is the Clebsch-Gordan coefficient (which depends upon the specificity of the $\ket{F, m_{\rm F}} \rightarrow \ket{F', m_{\rm F'}}$ transition and the polarization of the input beam driving the transition), and $L$ is the entire length of the cloud along $z$.

Next, we consider an atomic cloud held in an ODT with $N$ atoms, a fraction $F_{\rm BEC}$ of which have Bose condensed. In this case, the total atomic density $\rho (x,y,z)$ is the sum of thermal ($\rho_{\rm th}(x,y,z)$) and BEC ($\rho_{\rm B}(x,y,z)$) components. The thermal density distribution is Maxwell-Boltzmann
\begin{align}
\rho_{\rm th}(x,y,z) =~&~N_{\rm th}\left(\frac{m\omega_{\rm ho}^{2}}{2\pi k_{\rm B}T}\right)^{3/2} \nonumber \\ &\times \exp\left[-\frac{m}{2k_{\rm B}T}(\omega_{ x}^{2}x^{2} + \omega_{ y}^{2}y^{2} + \omega_{ z}^{2}z^{2}  )\right], \label{eqn16}  
\end{align}
where $N_{\rm th} = N(1 - F_{\rm BEC})$ is the number of thermal atoms, $T$ is the  temperature, \{$\omega_{\rm x},\omega_{\rm y},\omega_{\rm z}$\} are the trapping frequencies of the dipole potential along \{$x,y,z$\} and $\omega_{\rm ho} = (\omega_{ x}\omega_{ y}\omega_{ z})^{1/3}$ is their geometric mean. The characteristic size of the thermal component along a given direction is $\sigma_{i} = \sqrt{{k_{\rm B}T}/{m \omega_{\rm i}^{2}}}$.  

Similarly, the BEC distribution follows the so-called Thomas-Fermi profile
\begin{align}
\rho_{\rm B}(x,y,z) = \frac{\mu}{g}\left(1 - \frac{x^{2}}{R_{x}^{2}} - \frac{y^{2}}{R_{y}^{2}} - \frac{z^{2}}{R_{z}^{2}}\right), \label{eqn17}   
\end{align}
where $\mu / g$ is the peak density at the center of the condensate with $\mu$ being the chemical potential and $g = 4\pi\hbar^{2}a/m$ is the interaction parameter related to the s-wave scattering length $a$. The size of the condensate are specified by the Thomas-Fermi radii \{$R_{\rm x},R_{\rm y},R_{\rm z}$\}, where $R_{\rm i} = \sqrt{{2\mu}/{m \omega_{\rm i}^{2}}} $.

By integrating Eqs.~(\ref{eqn16}) and (\ref{eqn17}) along the beam propagation direction $z$, we find  the line-integraged  ``2D-densities''
\begin{align}
 \rho_{\rm th}^{\rm 2D}(x,y) &= \frac{N_{\rm th} m\omega ^{2}}{2\pi k_{\rm B}T}\exp\left[-\frac{m \omega^{2}}{2k_{\rm B}T}(x^{2} + y^{2})\right] \label{eqn18a}   
 \end{align}
 and
 \begin{align}
 \rho_{\rm B}^{\rm 2D}(x,y) &=  \frac{\mu}{g}\frac{4R_{\rm z}}{3}\Bigg[1 - \frac{x^{2} + y^{2}}{R_{\perp}^{2}} \Bigg]^{3/2}, \label{eqn18b}
\end{align}
where we have assumed $\omega_{\rm x} = \omega_{\rm y} \equiv \omega$ and  $R_{\rm x} = R_{\rm y} \equiv R_{\perp}$. Hence, the resonant optical depth for a transition on $^{87}$Rb-D1 line is
\begin{align}
& d_{\rm 0} = \frac{3\lambda^{2}\alpha^{2}}{2\pi}\left(\rho_{th}^{\rm 2D} + \rho_{B}^{\rm 2D}\right). \label{eqn19}     
\end{align}

As an example, for $^{87}$Rb D1-line: $J = J' = 1/2$ and the strongest transition is the $\ket{F=1, m_{\rm F} = 1} \rightarrow \ket{F' = 2, m_{\rm F'} = 2}$ driven by $\sigma^{+}$-polarized light (equivalently, $\ket{F=1, m_{\rm F} = -1} \rightarrow \ket{F' = 2, m_{\rm F'} = -2}$ driven by $\sigma^{-}$ polarization), with $\alpha^{\rm 2} = 1/2$.

Back-substituting Eq.~\ref{eqn19} into Eqs.~\ref{eqn14} and \ref{eqn12} gives us the effective optical depth $d$. Importantly, it is the effective optical depth that determines the maximum achievable memory efficiency. While the resonant (peak) optical depth $d_{0}$ is independent of the beam size and depends solely on the properties of atomic medium, $d$ depends upon the beam size as well. Beam sizes much larger than that of the cloud can result in a small $d$ and hence, smaller memory efficiency. Conversely, tightly focussed beams with beam diameters smaller than the cloud size would sample atoms mainly from the center of the trap (region of peak atomic density) making $d \approx d_{0}$ and thus, achieving the associated optimal efficiency.

\section*{Phase coherence of spin waves under a DC magnetic field}

\begin{figure*}[tb!]
\begin{center}
\includegraphics[width = 150mm]{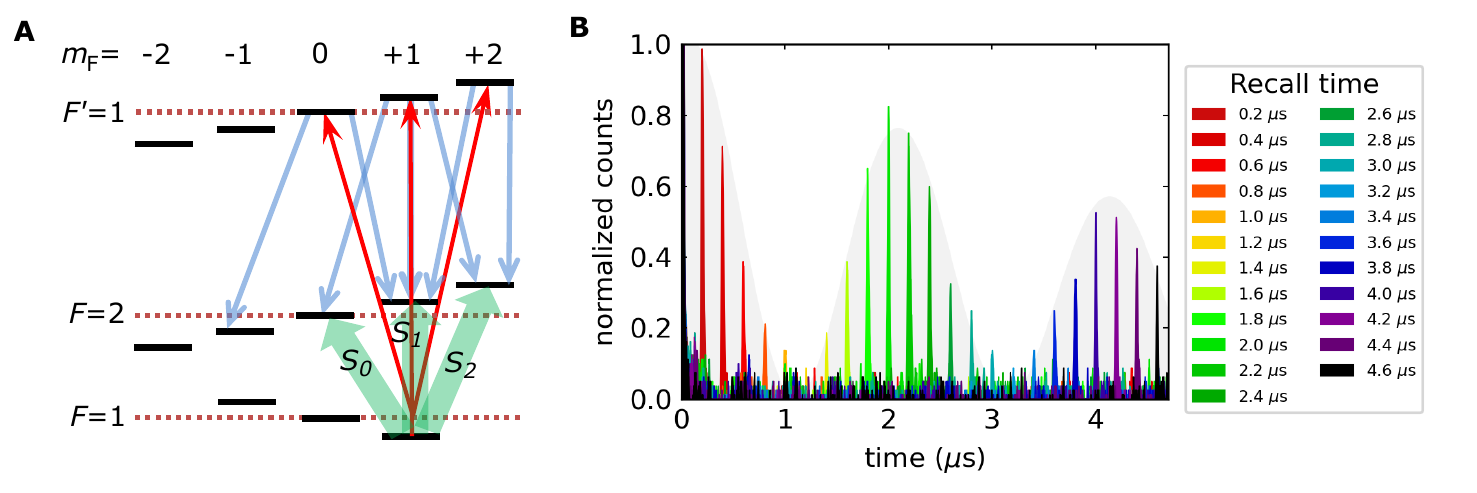} 
\caption{\textbf{Probing phase coherence during storage.}~\textbf{(A)}~Manipulation of stored spin excitations under an external DC magnetic field. The ground ($\ket{g}\equiv\ket{F= 1}$, $\ket{s}\equiv\ket{F= 2}$) and excited hyperfine ($\ket{e}\equiv\ket{F^{\prime}= 2}$) levels (dotted lines) are split into Zeeman sublevels with quantum numbers of $m_{F}$ and $m_{F^{\prime}}$, respectively. This allows the transfer of optical coherence onto three classes of spin excitations ($S_0$, $S_1$ and $S_2$) via various transition pathways for probe (red upward arrows) and control (blue downward arrows) fields, as further described in Methods. \textbf{(B)} Measured intensity of retrieved probe vs.\ storage time, measured for recall times from 0 to 4.4~$\mu$s in 200~ns intervals (i.e., each coloured peak in the histogram is from a different configuration with its own recall/storage time, and is associated with one of the black squares in Fig.~\ref{fig:single-photon}{\bf D}.}
\label{fig:coherence}
\end{center}
\end{figure*}

Without the components needed to create and store externally prepared quantum states (e.g. polarization qubits as in Refs.~\cite{Riedl2012,Vernaz-Gris2018}) in our current setup, we test whether phase coherence imparted by input probe light is preserved during storage by manipulating the phase evolution of the stored spin coherence, as depicted in Fig.~\ref{fig:BECdata}({\bf F}).  In addition to the bias magnetic fields used to cancel ambient fields ${\mathbf{B}_{\rm bias}\approx-\mathbf B}_{\rm amb}$), we apply a DC magnetic field ${\mathbf B}_{\rm ext}$  that splits the ground ($\ket{g}\equiv\ket{F= 1}$, $\ket{s}\equiv\ket{F= 2}$) and excited ($\ket{e}\equiv\ket{F^{\prime}= 2}$) hyperfine levels into Zeeman sublevels with quantum numbers of $m_{F}$ and $m_{F^{\prime}}$, respectively (Fig.~\ref{fig:coherence}({\bf A})). The energy spacing between the adjacent Zeeman sublevels is $\Delta_{F}=g_{F}\mu_{\rm B}|\mathbf{B}_{\rm ext}|$ and $\Delta_{F^{\prime}}=g_{F^{\prime}}\mu_{\rm B} |\mathbf{B}_{\rm ext}|$ in the ground ($F=1,2$) and excited ($F^{\prime}=2$) hyperfine levels, where $g_{F}$ ($g_{F^{\prime}}$) and $\mu_{\rm B}$ are Land\'e factor and Bohr magneton, respectively. 

In the absence of an external magnetic field, $\mathbf{B}_{\rm ext}=0$, the quantization axes are set by the propagation directions, both fields' polarizations are purely circular, and they drive $\sigma^{+}$ transitions only.  For $\mathbf{\rm B}_{\rm ext}\neq 0$, this external field sets the quantization axis, and both the probe and control fields may, in general, drive all $\sigma^{+}$-, $\sigma^{-}$- and $\pi$-transitions (depending on the orientations the quantization axis): the $\sigma^{+}$-component drives $\ket{F, m_F}\rightarrow\ket{F^{\prime}, m_F+1}$,  $\sigma^{-}$ drives $\ket{F, m_F}\rightarrow\ket{F^{\prime}, m_F-1}$ and $\pi$ drives and $\ket{F, m_F}\rightarrow\ket{F^{\prime}, m_F}$  (Fig.~\ref{fig:coherence}({\bf A})). In this scenario, spin-wave storage is achieved through multiple transition pathways, and optical coherence is mapped between the initially prepared ground state $\ket{F = 1, m_F = 1}$ and final states $\ket{F = 2, m_F = -1,0,1,2}$: in addition to the spin-wave  $S_{1}$ ($\ket{F=1,m_{F}=1}\rightarrow\ket{F=2,m_{F}=1}$), which dominates for $\vec{B}_{\rm ext}=0$,  new spin-waves denoted by $S_{0}$ ($\ket{F=1,m_{F}=1}\rightarrow\ket{F=2,m_{F}=0}$) and $S_{2}$ ($\ket{F=1,m_{F}=1}\rightarrow\ket{F=2,m_{F}=2}$) are formed (due to the weak transitions involved, the spin-wave involving $\ket{F=2,m_{F}=-1}$ is ignored from this point forward). The energy differences between the levels involved, $(E_{S_{0}}-E_{S_{1}})=-\tfrac{1}{2}\mu_{\rm B}|\mathbf{B}_{\rm ext}|\equiv -\hbar\omega$ and $(E_{S_{2}}-E_{S_{1}})=+\tfrac{1}{2}\mu_{\rm B}|\mathbf{B}_{\rm ext}|\equiv+\hbar\omega$, mean that  the spin excitations evolve with relative phases $e^{-i\omega t}$, $e^{+i\omega t}$  and $e^{0}$ for $S_{0}$, $S_{2}$ and $S_{1}$, respectively. As a result of interference among these spin-waves,  the intensity of retrieved photonic signals $I$ varies as $I\propto|q_{0}e^{-i\omega t}+ q_{1}+q_{2}e^{i\omega t}|^2$, where $q_0$, $q_1$ and $q_2$ are probability amplitudes associated with spin-waves $S_0$, $S_1$ and $S_2$, respectively. Thus, the intensity exhibits an oscillatory behavior~\cite{Matsukevich2006, Wang2011, Farrera2018} with a time period of 
\begin{align}
T = \frac{2\pi}{\omega}=\frac{2h}{\mu_{\rm B}|\mathbf{B}_{\rm ext}|},
\end{align}
as experimentally observed in Fig.~\ref{fig:coherence}({\bf B}).  

The degree of this interference (characterized by the visibility $V$) depends on the relative amplitudes of each spin-wave ($q_0, q_1$ and $q_2$), which are determined by both the amplitudes of the polarization components $\sigma^{+}$, $\sigma^{-}$ and $\pi$, and the respective transition matrix elements. While the weights of the polarization components belonging to probe and control beams are given by these beams' propagation-direction (Fig.~\ref{fig:setup}a) relative to $\mathbf{B}_{\rm ext}=({B_x},{B_y},{B_z})$, the transition probabilities are governed by by the selection rules and relevant Clebsch-Gordon coefficients. In our experiments, we achieve a maximum contrast for the interference visibility by optimizing $\mathbf{B}_{\rm ext}=({B_x= 0.8~{\rm G}},{B_y=0},{B_z=0})$, which nearly satisfies the condition of $q_1=2q_0=2q_2$ (and for the spin-wave involving $\ket{F=2,m_{F}=-1}$, with  $q_{-1}\approx0$). This results in an observation of an oscillation period of $T = 2.0~\mu \rm s$, reasonably close to the expected period of $T=1.7~\mu \rm s$.

Our measurements of visibility fall short of unity due to the deviation from the high-contrast interference condition ($q_1=2q_0=2q_2$) due to imperfect settings for polarization sates of the control and probe fields relative to quantization axis defined by the applied magnetic field. Furthermore, we observe a relatively small visibility ($V=62\%$) for single-photon-level measurements (the inset of Fig.~\ref{fig:BECdata}({\bf F})). We attribute this to the fact that the DC magnetic field is not constant over a 1-ms store-and-measure cycle (for multiple measurements) at which the atomic cloud is free falling in a distance comparable the size of the cloud itself. However, by performing single-shot measurements, which require a large input mean-photon number ($\overline{n}_{\rm in}\gg1$), we reach a visibility of $V=80\%$ that clearly reveals the phase-preserving character of the storage-and-recall process, as shown in Fig.~\ref{fig:single-photon}({\bf F}).

\end{document}